\documentclass{elsart}
\usepackage{graphics}
\usepackage{epsfig}
\usepackage{color}
\usepackage{enumerate}
\usepackage{ulem}

\def\qed{\hfill  \framebox(5,5){}}

\def\Res{{\rm Res}}

\expandafter\chardef\csname pre amssym.def at\endcsname=\the\catcode`\@ \catcode`\@=11

\def\undefine#1{\let#1\undefined}
\def\newsymbol#1#2#3#4#5{\let\next@\relax
 \ifnum#2=\@ne\let\next@\msafam@\else
 \ifnum#2=\tw@\let\next@\msbfam@\fi\fi
 \mathchardef#1="#3\next@#4#5}
\def\mathhexbox@#1#2#3{\relax
 \ifmmode\mathpalette{}{\m@th\mathchar"#1#2#3}%
 \else\leavevmode\hbox{$\m@th\mathchar"#1#2#3$}\fi}
\def\hexnumber@#1{\ifcase#1 0\or 1\or 2\or 3\or 4\or 5\or 6\or 7\or 8\or
 9\or A\or B\or C\or D\or E\or F\fi}

\font\tenmsa=msam10 \font\sevenmsa=msam7 \font\fivemsa=msam5
\newfam\msafam
\textfont\msafam=\tenmsa \scriptfont\msafam=\sevenmsa \scriptscriptfont\msafam=\fivemsa \edef\msafam@{\hexnumber@\msafam}
\mathchardef\dabar@"0\msafam@39
\def\dashrightarrow{\mathrel{\dabar@\dabar@\mathchar"0\msafam@4B}}
\def\dashleftarrow{\mathrel{\mathchar"0\msafam@4C\dabar@\dabar@}}

        \font\tenmsb=msbm10
\font\sevenmsb=msbm7 \font\fivemsb=msbm5
\newfam\msbfam
\textfont\msbfam=\tenmsb \scriptfont\msbfam=\sevenmsb \scriptscriptfont\msbfam=\fivemsb \edef\msbfam@{\hexnumber@\msbfam}
\def\Bbb#1{\fam\msbfam\relax#1}

\newtheorem{theorem}{{\bf Theorem}}
\newtheorem{remark}{{\bf Remark}}
\newtheorem{definition}[theorem]{{\bf Definition}}
\newtheorem{corollary}[theorem]{{\bf Corollary}}
\newtheorem{proposition}[theorem]{{\bf Proposition}}
\newtheorem{lemma}[theorem]{{\bf Lemma}}

\begin{document}
\begin{frontmatter}



\title{On the Shape of Curves that are Rational in Polar Coordinates}


\author[a]{Juan Gerardo Alc\'azar\thanksref{proy1}},
\ead{juange.alcazar@uah.es}
\author[b]{Gema Mar\'{\i}a D\'{\i}az-Toca\thanksref{proy2}},
\ead{gemadiaz@um.es}

\address[a]{Departamento de Matem\'aticas, Universidad de Alcal\'a,
E-28871 Madrid, Spain}
\address[b]{Departamento de Matem\'atica Aplicada, Universidad de
Murcia,  30100 Murcia, Spain}

\thanks[proy1]{Supported by the Spanish ``Ministerio de
Ciencia e Innovacion" under the Project MTM2011-25816-C02-01. Member of the Research Group {\sc asynacs} (Ref. {\sc ccee2011/r34})}

\thanks[proy2]{Supported by the Spanish ``Ministerio de
Ciencia e Innovacion" under the Project MTM2011-25816-C02-02.}


\begin{abstract}
In this paper we provide a computational approach to the shape
of curves which are rational in polar coordinates, i.e. which are defined by means of a parametrization
$(r(t),\theta(t))$ where both $r(t),\theta(t)$ are rational functions. Our study includes theoretical aspects on the shape of these curves, and algorithmic results which eventually lead to an algorithm for plotting the ``interesting parts" of the curve, i.e. the parts showing the main geometrical features.
\end{abstract}
\end{frontmatter}

\section{Introduction}\label{section-introduction}

Plotting and correct visualization of algebraic curves, both in the case when they are implicitly defined by means of a polynomial $f(x,y)=0$, or by a parametrization $\varphi(t)=(x(t),y(t))$ with $x(t),y(t)$ being rational, have received a great deal of attention in the literature on scientific computation (see for example \cite{JGDT}, \cite{Cheng}, \cite{Eigen}, \cite{Emel}, \cite{Lalo}, \cite{Hong}, \cite{seidel}). With this paper, we want to initiate a similar study for curves which are written in polar coordinates. Such curves may appear in Engineering and Physics, in particular in Mechanics (very specially in Celestial Mechanics), and also in Cartography. In this sense, here we address those curves which are rational when considered in polar form, i.e. curves defined by means of a parametrization $(r(t),\theta(t))$ where both $r(t),\theta(t)$ are rational functions. Arquimedes' spiral, Cote's spiral under certain conditions, Fermat's spiral or Lituus' spiral (some of these can be found in www.mathematische-basteleien.de/spiral.htm), for example, belong to this class; also, in Cartography these curves may appear when using for example conic, pseudo-conic or polyconic projections (\cite{Buga}, \cite{Sny}).

It is quite natural to start wondering if this kind of curves can be algebraic, when considered in cartesian coordinates. However, one may prove (see Theorem \ref{th-non-alg} in Subsection \ref{subsec-prelim}) that, with the exceptions of lines and circles, this cannot happen. As a consequence they can exhibit properties that algebraic curves cannot have. Here we analyze some of these which are relevant from the point of view of plotting, in particular the existence or not of infinitely many self-intersections (see Subsection \ref{subsec-self-int}), the appearance of limit points, limit circles and spiral branches, all of them introduced in Subsection \ref{subsec-limit}, and the relationship between both phenomena.

The possibility of detecting in advance the above phenomena is very useful to improve the plotting of these curves, which can be really ``devilish". To give an idea, one may use the computer algebra system Maple, where the instruction {\tt polarplot} is available (we have also tried other packages like Maxima, SAGE and Mathematica; however, either no similar instruction was available, or the corresponding command behaved in a very similar way to {\tt polarplot}). This instruction allows to draw curves defined in polar coordinates either by means of an equation $r=f(\theta)$, or by a parametrization $(r(t),\theta(t))$, and has a nice performance for simple curves, but may not be enough for illustrating the behavior of a more complicated curve. As an example, one may consider the following parametrizations in polar coordinates: (1) $r=\displaystyle{\frac{t^2}{t^2-11t+30}}, \theta=\displaystyle{\frac{t^2+78}{t^2+1}}$; (2) $r=t, \theta=\displaystyle{\frac{t^2+14}{t^2+1}}$; (3) $r=t, \theta=\displaystyle{\frac{t^3+1}{t^2-3t+2}}$. If one uses {\tt polarplot} to visualize these curves, one obtains the outputs in Fig. 1: here the curves (1), (2), (3) are displayed from left to right; in (1) and (2) we have asked Maple to plot the curve for $t\in (-\infty,\infty)$, and in the case of (3) we have chosen $t\in (-2.01,2.01)$ (because $r,\theta$ are both non-bounded for $t\to \pm \infty$). However, in the three cases it is clear that the output does not really help to understand the behavior of the curve.

\begin{figure}[ht]
\begin{center}
\centerline{$\begin{array}{ccc}   \begin{array}{c}\psfig{figure=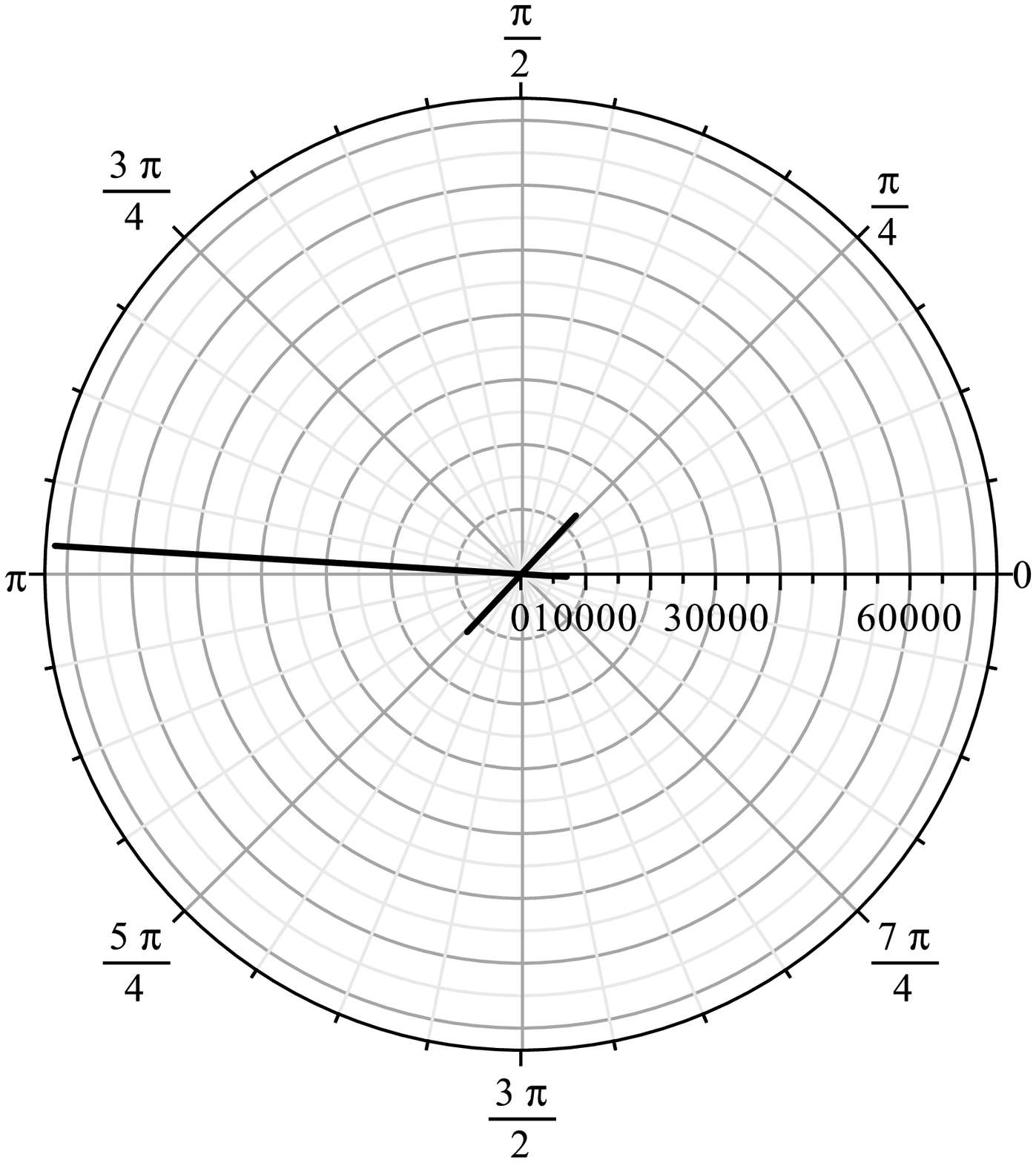,width=5cm,height=5cm} \end{array}   &
 \begin{array}{c} \psfig{figure=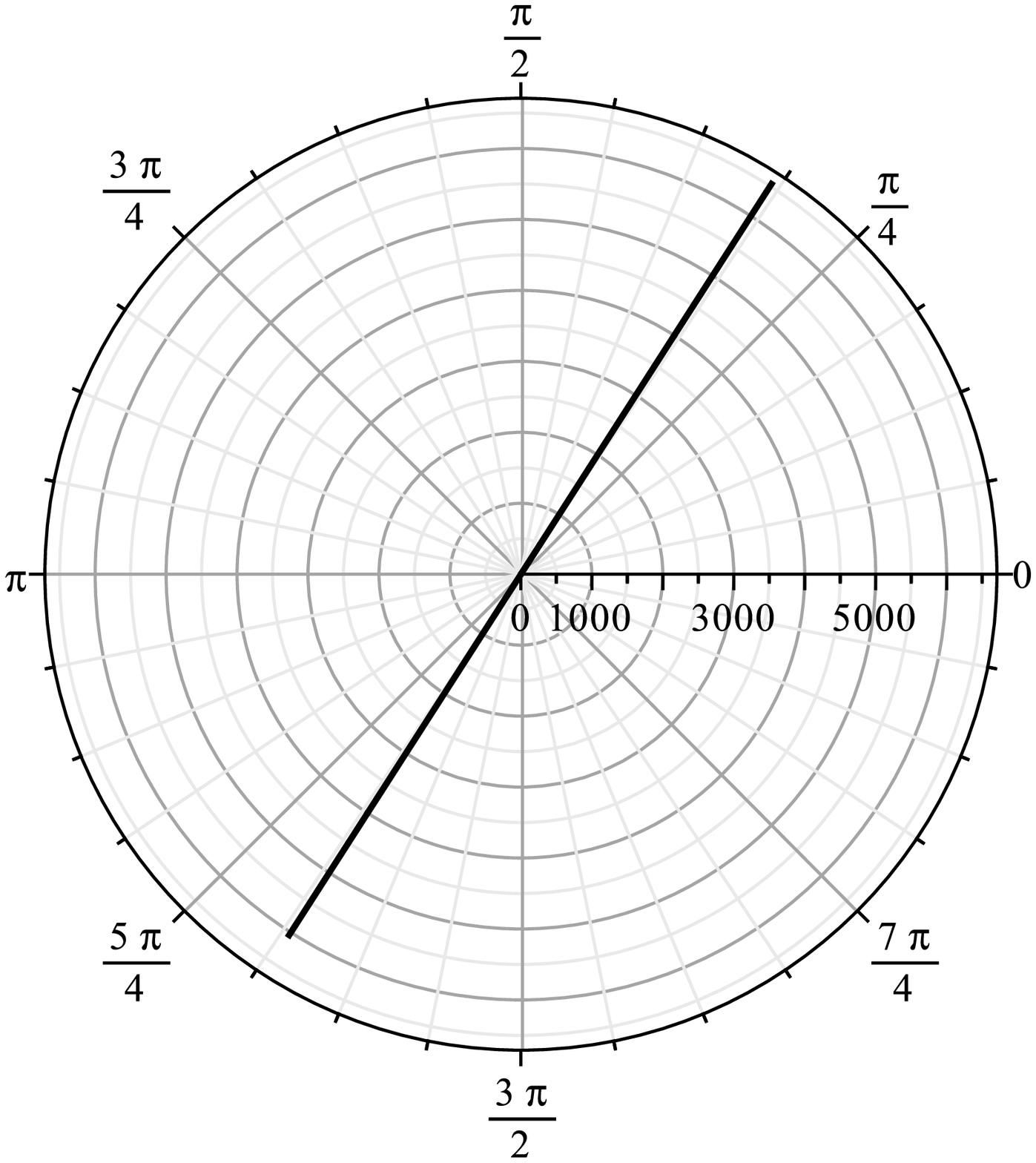,width=5cm,height=5cm}\end{array} & \begin{array}{c}\psfig{figure=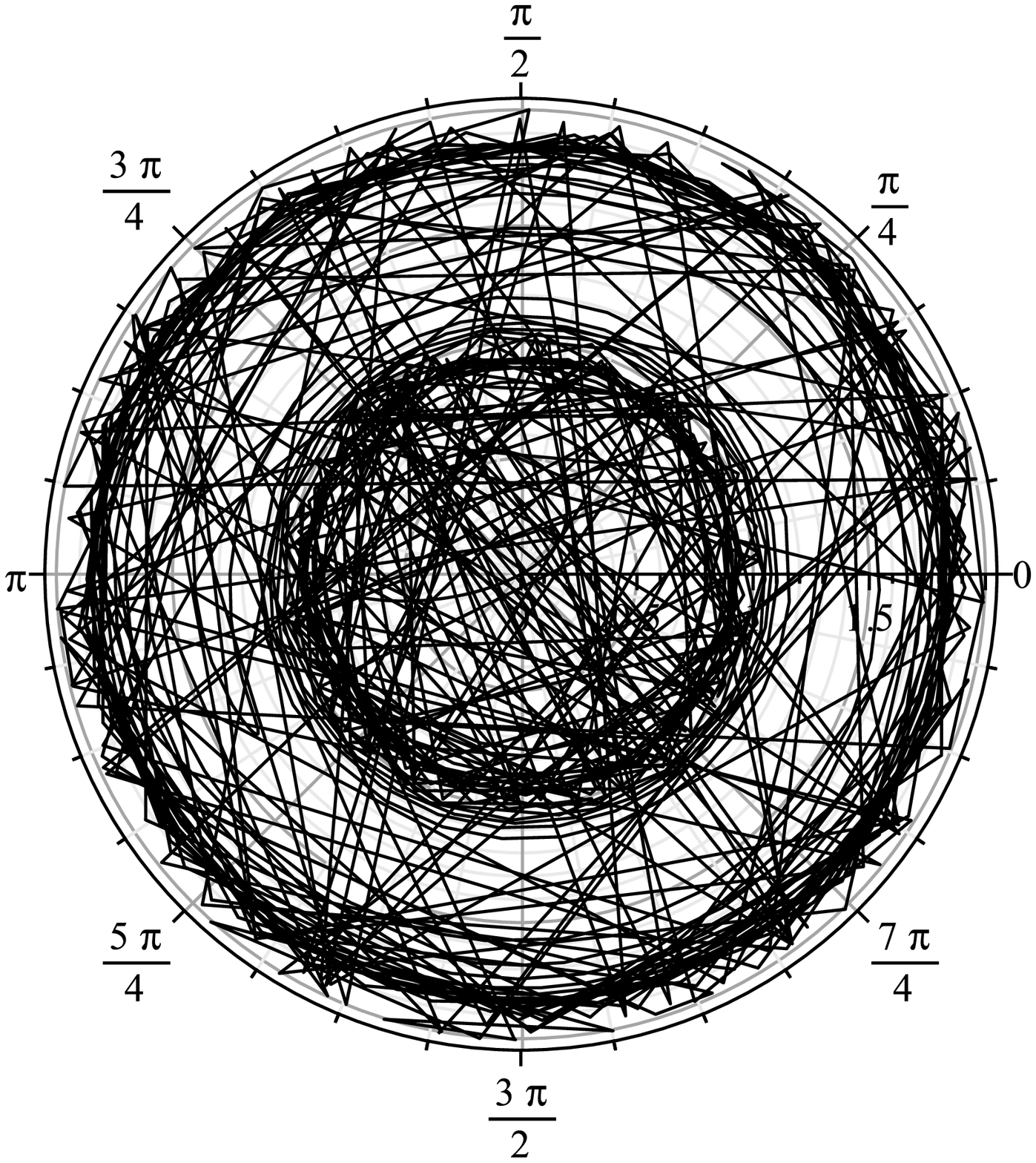,width=5cm,height=5cm}  \end{array}
\end{array}$}
\end{center}
\caption{Some complicated curves plotted with {\tt polarplot}}
\end{figure}

By using our ideas, one can obtain an algorithm which provides, for a given curve of the considered kind, information and several plottings corresponding to the more interesting parts of it. We have implemented this algorithm in Maple 15. Our algorithm analyzes the curve, detects its main features, and uses the command {\tt polarplot} for plotting the curve over several intervals (generated by our algorithm), each one showing certain features of the curve; together with the provided information, these plottings help to clarify the behavior of the curve. The outputs of our algorithm for curves in Fig. 1 can be checked in Subsection 3.2  and Subsection 3.4.

The paper is structured as follows: in Section 2, the reader will find the theoretical results on the shape of these curves. In Section 3, we provide some details on the algorithm, together with several examples of outputs. Finally, in Section 4 we present some conclusions and suggest future lines of research.

\section{Geometrical Properties of Curves Rational in Polar Coordinates}\label{sec-background}

\subsection{Preliminaries and First Properties}\label{subsec-prelim}

Along the paper, given a point $P\in {\Bbb R}^2$ we denote its polar coordinates by $(r,\theta)$, and its cartesian coordinates by $(x,y)$.
Notice that the polar coordinates of $P$ are not unique, since $(r,\theta)$, $(r,\theta+2k\pi)$ with $k\in {\Bbb Z}$, or
$(-r,\theta+(2\ell+1)\pi)$ with $\ell \in {\Bbb Z}$, define the same point.

We will analyze the geometry of a planar curve
${\mathcal C}$ which is {\it rational} when parametrized in polar coordinates. In other words, there exist four real
polynomials $A(t)$, $B(t)$, $C(t)$, $D(t)$ not all of them constant, with $\gcd(A,B)=\gcd(C,D)=1$, such that
\[
\varphi(t)=(r(t),\theta(t))=\left(\displaystyle{\frac{A(t)}{B(t)},\frac{C(t)}{D(t)}}\right), \,\, t \in {\Bbb R},
\]
parametrizes $\mathcal C$ in polar coordinates. We will also consider the curve ${\mathcal C}^{\star}=\varphi({\Bbb C})$, i.e. the rational planar curve parametrized by $\varphi(t)$ over the $(r,\theta)$--plane; for example, if $\varphi(t)=(t,t)$ (which parametrizes an Arquimedes spiral, then ${\mathcal C}^{\star}$ is a line over the $(r,\theta)$ plane. Notice that our goal is not to describe ${\mathcal C}^{\star}$, but still this curve is useful at several steps for proving certain facts on ${\mathcal C}$. 

Furthermore we assume that $\varphi(t)$, as a parametrization of ${\mathcal C}^{\star}$, is {\it proper}, i.e. that it is
injective for almost all values of $t$
(equivalently, that almost all points of ${\mathcal C}^{\star}$ are generated by just one $t$-value). Observe that if $\varphi(t)$ is not proper then it can always be properly reparametrized by applying the algorithm
in Chapter 6 of \cite{SWPD}. Also, we will say that a point $P_0\in {\mathcal C}^{\star}$ is {\it reached} by the parametrization if there exists some
$t_0\in {\Bbb C}$ such that $\varphi(t_0)=P_0$. It can be proven (see Proposition 4.2 in \cite{Andradas})
 that the only point of ${\mathcal C}^{\star}$ that may not be reached by the parametrization is

\[P_{\infty}=\mbox{lim}_{t\to \pm \infty}\varphi(t)\]

whenever it exists. Furthermore, if $\varphi(t)$ is proper then the real points of ${\mathcal C}^{\star}$ that are reached by complex, but not real, values of the
parameter are, except perhaps for $P_{\infty}$, isolated points of ${\mathcal C}^{\star}$ (see also Proposition 4.2 in \cite{Andradas}). Therefore, the
real part of ${\mathcal C}^{\star}$ can be expressed as
\[
\varphi({\Bbb R})\cup P_{\infty}\cup \{\mbox{Finitely many isolated points}.\}
\]

In the sequel we will discard the isolated points of ${\mathcal C}^{\star}$ and we will focus on the remaining part of ${\mathcal C}^{\star}$. Since polar coordinates provide a natural mapping $\Pi$ (not 1:1) from ${\mathcal C}^{\star}$ onto ${\mathcal C}$, we will say that a point of ${\mathcal C}$ is {\it reached} by $\varphi(t)$ if it has the form $\Pi(P)$, where $P\in {\mathcal C}^{\star}$ has been reached by $\varphi(t)$. In this language, notice that we are identifying ${\mathcal C}=\Pi(\tilde{\mathcal C}^{\star})$ where $\tilde{\mathcal C}^{\star}$ is the real part of ${\mathcal C}^{\star}$, discarding isolated singularities.

 Now if ${\mathcal C}$ is either a circle centered at the origin or a line passing through the origin, then it is algebraic in algebraic and polar form at the same time (in fact, ${\mathcal C}^{\star}$ is represented by an equation of the type $r-r_0=0$ or $\theta-\theta_0=0$). The next theorem proves that circles and lines are the only cases when this phenomenon happens. We acknowledge here the help of Fernando San Segundo for proving the theorem. Here we use the expression {\it real curve} to mean a curve with infinitely many real points.


\begin{theorem} \label{th-non-alg}  Let ${\mathcal C}$ be a real algebraic curve. Then it is rational in polar coordinates if and only if
it is either a real circle centered at the origin or a real line passing through the origin.
\end{theorem}

 {\bf Proof.}  The implication $(\Leftarrow)$ is clear; so, let us focus on $(\Rightarrow)$. For this purpose, let $f(r,\theta)\in {\Bbb R}[r,\theta]$ be the implicit equation of ${\mathcal C}^{\star}$. Now since ${\mathcal C}^{\star}$ is rational, then it is irreducible, and so also $f(r,\theta)$. If $f$ depends only on $r$, taking into account that $f$ is irreducible and that
 ${\mathcal C}^{\star}$ is a real curve (because $r(t),\theta(t)$ are real functions), $f$ must be linear. So, ${\mathcal C}$ can be either a circle or a point (the origin, if $f(r,\theta)=r$); but in this last case ${\mathcal C}$ is not a real curve, and hence it must be a circle. If $f$ depends only on $\theta$, then for similar reasons ${\mathcal C}$ must be a line. Thus, assume that $f$ depends on both $r,\theta$, and let us see that in this case ${\mathcal C}$ cannot be algebraic. Indeed, if ${\mathcal C}$ is algebraic, let $g(x,y)\in {\Bbb R}[x,y]$ be its implicit equation and let $h(r,\theta)=g(r\mbox{cos}(\theta),r\mbox{sin}(\theta))$. Notice that every zero of $f(r,\theta)$ is also a zero of $h(r,\theta)$.
In this situation, consider
the resultant $M(\theta,u,v)=\Res_r(f(r,\theta),h(r,u,v))$, where $h(r,u,v)$ is the function obtained when substituting formally
$\mbox{sin}(\theta)=u$ and $\mbox{cos}(\theta)=v$ in $h(r,\theta)$.  Since ${\mathcal C}^{\star}$ is irreducible and $f$
depends on both $r,\theta$, does not have any factor only depending on $r$, and therefore $f(r,\theta)$ and $h(r,u,v)$ cannot
have any factor in common; hence, $M(\theta,u,v)$ cannot be identically $0$.
Observe also that $M(\theta,u,v)$ must depend explicitly on $\theta$, i.e. $M\neq M(u,v)$. Now by Lemma 4.3.1 in \cite{Wi96} we have that the resultant $\Res_r(f(r,\theta),h(r,u,v))$ specializes properly when
$u=\mbox{sin}(\theta)$ and $v=\mbox{cos}(\theta)$, i.e.
\[M(\theta,\mbox{sin}(\theta),\mbox{cos}(\theta))=\Res_r(f(r,\theta),h(r,\mbox{sin}(\theta),\mbox{cos}(\theta))).\]
Since $f$ depends on both $r,\theta$ we can find an open interval $I\subset {\Bbb R}$ such that for every $\theta_0\in I$, the equation $f(r,\theta_0)=0$ and therefore also $h(r,\theta_0)=0$, have at least one real solution. Hence, by well-known properties of resultants, for all $\theta_0\in I$ it
holds that $\theta=\theta_0$ is a zero of $M(\theta,\mbox{sin}(\theta),\mbox{cos}(\theta))$, i.e. $M(\theta,\mbox{sin}(\theta),\mbox{cos}(\theta))$ vanishes over $I$. Hence, by the
Identity Theorem (see page 81 in \cite{Jong}), and taking into account that $M$ is analytic, it holds that $M(\theta,\mbox{sin}(\theta),\mbox{cos}(\theta))$ is identically $0$. Since $M$ is a polynomial in
$\theta$, $\mbox{sin}(\theta)$, $\mbox{cos}(\theta)$ where $\theta$ is present (i.e. $M\neq M(\mbox{sin}(\theta),\mbox{cos}(\theta))$), this implies that $\theta$, $\mbox{sin}(\theta)$, $\mbox{cos}(\theta)$, are algebraically dependent. However, this cannot happen
because $\mbox{sin}(\theta)$, $\mbox{cos}(\theta)$ are trascendental functions. \qed

Hereafter we exclude the cases when either $r(t)$ or $\theta(t)$ are constant.

\subsection{Self-intersections}\label{subsec-self-int}

A point generated by $t\in {\Bbb R}$ corresponds to a self-intersection of
${\mathcal C}$ if it exists $s\in {\Bbb R}$, $t\neq s$, and $k\in {\Bbb Z}$, such that $(t,s,k)$ is solution of some of the following two
systems:
\[
\begin{array}{cc}
(\star)_1=\left\{\begin{array}{l} r(t)=r(s)\\
\theta(t)=\theta(s)+2k\pi\end{array}\right. & (\star)_2=\left\{\begin{array}{l} r(t)=-r(s)\\
\theta(t)=\theta(s)+(2k+1)\pi\end{array}\right.\end{array}
\]
Moreover, when $P_{\infty}=(r_{\infty},\theta_{\infty})$ exists, we also have to consider the self-intersections involving
$P_{\infty}$, i.e. the solutions of
\[
\begin{array}{cc}
(\star)_3=\left\{\begin{array}{l} r_{\infty}=r(s)\\
\theta_{\infty}=\theta(s)+2k\pi\end{array}\right. & (\star)_4=\left\{\begin{array}{l} r_{\infty}=-r(s)\\
\theta_{\infty}=\theta(s)+(2k+1)\pi\end{array}\right.\end{array}\]
So, in order to study the self-intersections of ${\mathcal C}$ we have to study the
above systems. Additionally, whenever the equation $r(t)=0$ has more than one solution, or one solution and $r_{\infty}=0$, the origin is also a self-intersection.

In the algebraic case, Bezout's theorem forces every algebraic curve to have
finitely many self-intersections. However, in our case this does not necessarily hold.  Since $r(t)$ is rational, it cannot happen that a same point of ${\mathcal C}$ is crossed by infinitely many branches of ${\mathcal C}$. But there can be infinitely many {\it different} self-intersections, which can be detected by analyzing $(\star)_1$ and $(\star)_2$. We start with $(\star)_1$. For this purpose, we write


\[
\begin{array}{l}
r(t)-r(s)=\displaystyle{\frac{A(t)B(s)-A(s)B(t)}{B(t)B(s)}},\\ \\
\theta(t)-\theta(s)-2k\pi=\displaystyle{\frac{C(t)D(s)-C(s)D(t)-2k\pi D(t)D(s)}{D(t)D(s)}}.
\end{array}
\]
We denote the numerator of $r(t)-r(s)$ by $\alpha(t,s)$, and the numerator of $\theta(t)-\theta(s)-2k\pi$ by $\beta(t,s,k)$.

\begin{lemma} \label{lem-inst-1}
There do not exist $a(t,s)\neq 1$ and $k_0\in {\Bbb Z}$ (in fact, $k_0\in {\Bbb R}$) such that $a(t,s)$ simultaneously divides $\gcd(\alpha(t,s),\beta(t,s,k_0))$ and
$B(t)\cdot B(s)\cdot D(t) \cdot D(s)$.
\end{lemma}

{\bf Proof.} Assume that  we have an irreducible polynomial $a(t,s)$ and $k_0\in {\Bbb Z}$ such that $a(t,s)$ divides both $\gcd(\alpha(t,s),\beta(t,s,k_0))$ and $B(t)\cdot B(s)\cdot D(t) \cdot D(s)$.  We need to show that $a(t,s)=1$. Suppose that $a(t,s)$ divides $B(t)$ (resp. $B(s)$). Since it also divides $\gcd(\alpha(t,s),\beta(t,s,k_0))$, it divides $\alpha(t,s)$, and therefore it also divides $A(t)$ (resp. $A(s)$).
On the other hand, $\gcd(A(t),B(t))=1$ by hypothesis. That shows that $a(t,s)=1$.
If $a(t,s)$ divides $D(t)$ (resp. $D(s)$) we argue similarly with $\beta(t,s,k)$. \qed

Then we are ready to proceed with the following theorem, that will have important consequences on the study of $(\star)_1$.

\begin{theorem} \label{th-1}
 For any $k\neq 0$, $k\in {\Bbb Z}$, (in fact, $k\in {\Bbb R}$) the system $(\star)_1$ has finitely many solutions.
\end{theorem}

{\bf Proof.}  In order to prove the statement, we need to show that for $k\in {\Bbb Z}$, $k\neq 0$, $\gcd(\alpha(t,s),\beta(t,s,k))=1$.
 For this purpose, assume by contradiction
that there exists some $k\in {\Bbb Z}$, $k\neq 0$, such that $H(t,s)=\gcd(\alpha(t,s),\beta(t,s,k))\neq 1$. Then $H(t,s)$ defines an algebraic curve ${\mathcal H}$ over ${\Bbb C}^2$; furthermore, since by hypothesis $k\neq 0$, $H(t,s)$ cannot be $t-s$. So, there are infinitely many points $(t_0,s_0)\in {\mathcal H}$ with $t_0\neq s_0$. Now by Lemma \ref{lem-inst-1} only finitely many of them fulfill $B(t)\cdot B(s)\cdot D(t) \cdot
D(s)=0$. So, the functions $r(t)$, $r(s)$, $\theta(t)$ and
$\theta(s)$ are well-defined at almost all the points $(t_0,s_0)\in {\mathcal H}$, and $r(t_0)=r(s_0)$, $\theta(t_0)=\theta(s_0)+2k\pi$. Since for a fixed $t_0$ there can only be finitely many values
of $s$ such that $r(t_0)=r(s)$ (because $r(t)$ is rational) we deduce then that there are infinitely many points $(r_0,\theta_0)\in {\mathcal C}^{\star}$
such that $(r_0,\theta_0+2k\pi)\in {\mathcal C}^{\star}$ too. In other words, the curves defined over the $(r,\theta)$-plane by $f(r,\theta)$ and $f(r,\theta+2k\pi)$ have infinitely many points in common, and since $f$ is irreducible and both have the same degree, then they must
define ${\mathcal C}^{\star}$. Now let $(r_0,\theta_0)\in {\mathcal C}^{\star}$; then
$(r_0,\theta_0+2k\pi)$ is a zero of $f(r,\theta+2k\pi)$, and since $f(r,\theta+2k\pi)$ also defines ${\mathcal C}^{\star}$ then it
is also a point of ${\mathcal C}^{\star}$. Following the same reasoning, we conclude that
$(r_0,\theta_0+4k\pi)$ also belongs to ${\mathcal C}^{\star}$, and in fact that $(r_0,\theta_0+2nk\pi)\in {\mathcal C}^{\star}$ for all $n\in {\Bbb
N}$. Since $k\neq 0$ we have that these are all different points of ${\mathcal C}^{\star}$. Hence, we have that ${\mathcal C}^{\star}$ intersects the line $r=r_0$ at infinitely many points. But this is impossible because ${\mathcal C}^{\star}$ is
algebraic. \qed

We can obtain similar results for $(\star)_2$. For this purpose, we denote the numerator of $r(t)+r(s)$ by $\mu(t,s)$, and the numerator of
$\theta(t)-\theta(s)-(2k+1)\pi$ by $\nu(t,s,k)$. Then the following lemma, analogous to Lemma \ref{lem-inst-1}, holds.

\begin{lemma} \label{lem-inst-2}
There does not exist $b(t,s)\neq 1$ and $k_0\in {\Bbb Z}$ (in fact, $k_0\in {\Bbb R}$) such that $b(t,s)$ simultaneously divides $\gcd(\mu(t,s),\nu(t,s,k_0))$ and $B(t)\cdot
B(s)\cdot D(t) \cdot D(s)$.
\end{lemma}

The following theorem, very similar to Theorem \ref{th-1}, holds.

\begin{theorem} \label{th-2}
 For any $k\in {\Bbb Z}$ (in fact, $k_0\in {\Bbb R}$), it holds that $\gcd(\mu(t,s),\nu(t,s,k))= 1$.
\end{theorem}

{\bf Proof.} Arguing by contradiction as in the proof of Theorem \ref{th-1}, we conclude that $f(r,\theta)$ and $f(-r,\theta+2k\pi)$ define the
same curve (namely, ${\mathcal C}^{\star}$). So, let $(r_0,\theta_0)\in {\mathcal C}^{\star}$ where $r-r_0$ does not divide $f(r,\theta)$. Then
$(-r_0,\theta_0+(2k+1)\pi)\in {\mathcal C}^{\star}$, and for the same reason $(r_0,\theta_0+(4k+2)\pi)\in {\mathcal C}^{\star}$ too. Proceeding
this way we get that all the points of the form $(r_0,2n(k+1)\pi)$, with $n\in {\Bbb N}$, belong to  ${\mathcal C}^{\star}$. For any value $k\in
{\Bbb Z}$ all these points are different; so, we get that the intersection of $ {\mathcal C}^{\star}$ with the line $r=r_0$ consists of infinitely
many different points. But this cannot happen because $ {\mathcal C}^{\star}$ is algebraic. \qed

Hence, by Theorem \ref{th-1} and Theorem \ref{th-2}, we obtain the following result on the existence of infinitely many
self-intersections of ${\mathcal C}$.
\begin{corollary} \label{corol-3}
${\mathcal C}$ has infinitely many self-intersections if and only if $(\star)_1$ or $(\star)_2$ have solutions for infinitely many values $k\in
{\Bbb Z}$.
\end{corollary}

Based on Corollary \ref{corol-3}, we have the following result which provides a sufficient condition for ${\mathcal C}$ to have finitely many
self-intersections.

\begin{theorem} \label{th-3}
If $\theta(t)$ is bounded, then there are at most finitely many self-intersections.
\end{theorem}

{\bf Proof.}  If $\theta(t)$ is bounded, the number of integer values of $k$ satisfying the second equation of $(\star)_1$ or $(\star)_2$ for some $(t,s)$ is necessarily finite. Then the result follows from Corollary \ref{corol-3}.  \qed

The converse of Theorem \ref{th-3} is not necessarily true (see Example 2). So, we still need a characterization for the existence of infinitely
many self-intersections.
\begin{lemma} \label{need}
The polynomial $\Res_s(\alpha(t,s),\beta(t,s,k))$ cannot be constant. More precisely, it has positive degree in $k$.\end{lemma}

{\bf Proof.}  By Theorem \ref{th-1}, $\Res_s(\alpha(t,s),\beta(t,s,k))$ cannot be identically $0$. Moreover, by writing explicitly the system ${\mathcal S}\equiv \{\alpha(t,s)=0,\beta(t,s,k)=0\}$,
\[
\left\{\begin{array}{l}
A(t)B(s)-A(s)B(t)=0\\
C(t)D(s)-C(s)D(t)-2k\pi D(t)D(s)=0
\end{array}\right.
\]
we observe that the points $(t',t',0)$, $t' \in \Bbb{R}$, are solutions of ${\mathcal S}$. So, for any $t'\in\Bbb{R}$, $\Res_s(\alpha(t,s),\beta(t,s,k))$ vanishes at $(t',0)$ and therefore it cannot be constant. For the same reason,  $\Res_s(\alpha(t,s),\beta(t,s,k))$ cannot be an univariate polynomial in $t$. In fact, $k$ is a divisor of  $\Res_s(\alpha(t,s),\beta(t,s,k))$. \qed

\begin{lemma} \label{need2}
The polynomial $\Res_s(\mu(t,s),\nu(t,s,k))$ cannot be constant. Moreover,  if $(\star_2)$ has solutions for infinitely many values of $k\in {\Bbb Z}$, then $\Res_s(\mu(t,s),\nu(t,s,k))$ has positive degree in $k$.
\end{lemma}

{\bf Proof.} By Theorem \ref{th-2}, $\Res_s(\mu(t,s),\nu(t,s,k))$ cannot be identically $0$. Moreover, by writing explicitly the system ${\mathcal S}'\equiv \{\mu(t,s)=0,\nu(t,s,k)=0\}$,
\[
\left\{\begin{array}{l}
A(t)B(s)+A(s)B(t)=0\\
C(t)D(s)-C(s)D(t)-(2k+1)\pi D(t)D(s)=0
\end{array}\right.
\]
we observe that if $A(t')=0$ with $D(t')\neq 0$ then there exists $k'\in \Bbb{R}$ such that the point $(t',t',-1/2)$ is a solution of ${\mathcal S}$. If $D(t') =0$, then the points $(t',t',k)$ are solutions of ${\mathcal S}$ for any $k$. As a consequence,   the polynomial $\Res_s(\mu(t,s),\nu(t,s,k))$ cannot be a constant.

Now assume that $(\star)_2$ has infinitely many solutions. That implies that there are infinitely $(t',s',k')$  solutions of ${\mathcal S}$ with $D(t')\neq 0$, $D(s')\neq 0$, $B(t')\neq 0$ and  $B(s')\neq 0$ . Due to Theorem \ref{th-2} and the linearity of $k$ in $\nu(t,s,k)$,
the polynomial $\Res_s(\mu(t,s),\nu(t,s,k))$  has positive degree in both $t$ and $k$.
\qed

Next we denote by $\xi_1(t,k)$ (resp. $\xi_2(t,k)$) the result of taking out
from the square-free part of $\Res_s(\alpha(t,s),\beta(t,s,k))$ (resp. $\Res_s(\mu(t,s),\nu(t,s,k))$) the univariate factors in
$t$.

\begin{theorem} \label{th-4}
 ${\mathcal C}$ has infinitely many self-intersections if and only if either $\xi_1(t,k)=0$ or $\xi_2(t,k)=0$
is an algebraic curve (in the
$\{t,k\}$-plane), non-bounded in $k$.
\end{theorem}

{\bf Proof.}
If ${\mathcal C}$ has infinitely many self-intersections,  by Corollary \ref{corol-3} one of the systems $(\star)_1$, $(\star)_2$ has solutions for infinitely many values $k\in {\Bbb Z}$. By reasoning over $(\star)_1$, $(\star)_2$, one may see that each value $t=t_0$ or $s=s_0$ gives rise to just finitely many solutions of the system (otherwise, we would reach the conclusion that $\gcd(C,D)\neq 1$). Consequently, if some system $(\star)_1$, $(\star)_2$ has solutions for infinitely many values $k\in {\Bbb Z}$, ${\mathcal S}\equiv \{\alpha(t,s)=0,\beta(t,s,k)=0\}$ or ${\mathcal S'}\equiv \{\mu(t,s)=0,\nu(t,s,k)=0\}$ has infinitely many solutions too. Since resultants are  combinations of the polynomials which define the systems ${\mathcal S}$ and ${\mathcal S'}$ and they are non-zero by Lemma \ref{need} and Lemma \ref{need2}, either $\xi_1(t,k)=0$ or $\xi_2(t,k)=0$ is non-bounded in $k$.

Conversely,  assume that $\xi_1(t,k)$ is
non-bounded in $k$ (we would argue in a similar way with $\xi_2(t,k)=0$). Since it has by definition no univariate factors depending on $t$, there are at most finitely many points $(t',k')$ with $\xi_1(t',k')=0$ where the leading coefficient of $\alpha(t,s)$ with respect to $s$, $D(t)$ and $B(t)$ vanish. By the Extension Theorem of resultants, any
other point of $\xi_1(t,k)=0$ corresponds to a solution of $(\star)_1$. Since $\xi_1(t,k)=0$ is non-bounded in $k$, observe that it contains infinitely many points with $k\in {\Bbb Z}$ and so with $t\neq s$. Therefore  ${\mathcal C}$ has infinitely many self-intersections. \qed

The above results are illustrated in the following examples.

\begin{figure}[ht]
\begin{center}
\centerline{ \psfig{figure=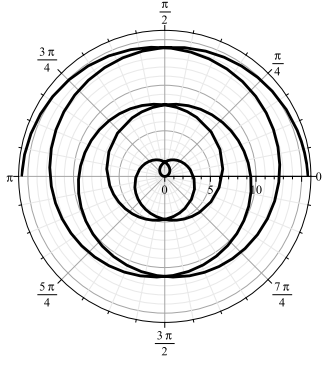,width=5cm,height=5cm} }
\end{center}
\caption{$\varphi(t)=(t,t)$: $\theta(t)$ non-bounded, infinitely many self-intersections}\label{ej1}
\end{figure}

{\bf Example 1.} Let ${\mathcal C}$ be parametrized in polar coordinates by $\varphi(t)=(t,t)$. The plotting of this curve for $t\in
[-5\pi,5\pi]$ is shown in Figure \ref{ej1}. One may see that $\theta(t)=t$ is not bounded; so, there might be infinitely many self-intersections. In this case we get $\xi_1(t,k)=2k\pi$ and $\xi_2(t,k)=2t-2k\pi-\pi$.
Finally, it is easy to see that $\xi_1(t,k)=0$ is bounded in $k$, but $\xi_2(t,k)=0$ is not. So, from Theorem \ref{th-4} we conclude that
${\mathcal C}$ has infinitely many self-intersections. In fact, one may see that all these self-intersections lay on the $y$-axis.

\begin{figure}[ht]
\begin{center}
\centerline{ \psfig{figure=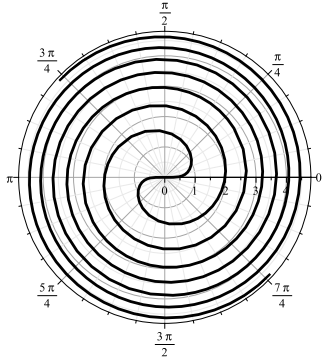,width=5cm,height=5cm} }
\end{center}
\caption{$\varphi(t)=\left(t,\displaystyle{\frac{t^4}{t^2+1}}\right)$: $\theta(t)$ non-bounded, finitely many self-intersections}\label{ej2}
\end{figure}

{\bf Example 2.} Let ${\mathcal C}$ be parametrized in polar coordinates by $\varphi(t)=\left(t,\displaystyle{\frac{t^4}{t^2+1}}\right)$. Again
$\theta(t)$ is not bounded. However, the plotting of this curve for $t\in \left[\displaystyle{-\frac{3}{2},\frac{3}{2}}\right]$, shown in Figure
\ref{ej2}, suggests that the curve has not infinitely many self-intersections (in fact, it has no self-intersections at all). Let us check this by using
Theorem \ref{th-4}. We get
$\xi_1(t,k)=2k\pi$ and $\xi_2(t,k)=(2k+1)\pi$. Both curves are clearly bounded in $k$; therefore, there are at most finitely many
self-intersections.

\begin{figure}[ht]
\begin{center}
\centerline{ \psfig{figure=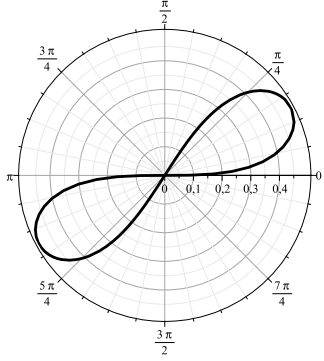,width=5cm,height=5cm} }
\end{center}
\caption{$\varphi(t)=\left(\displaystyle{\frac{t}{t^2+1},\frac{t^2}{t^2+1}}\right)$: $\theta(t)$ bounded.}\label{ej3}
\end{figure}

{\bf Example 3.} Let ${\mathcal C}$ be parametrized by $\varphi(t)=\left(\displaystyle{\frac{t}{t^2+1},\frac{t^2}{t^2+1}}\right)$. Since
$\theta(t)=\displaystyle{\frac{t^2}{t^2+1}}$ is bounded, from Theorem \ref{th-3} it follows that there are at most finitely many
self-intersections. Furthermore,  $\xi_1(t,k)=2k\pi(2k \pi t^2+2k\pi-t^2+1)$ and
$\xi_2(t,k)=\pi(2k+1)(2k\pi t^2+2k \pi-t^2+\pi t^2+1+\pi)$, both bounded in $k$. So, by Theorem \ref{th-4} we derive the same
conclusion. The plotting of the curve for $t\in (-\infty,\infty)$ is shown in Figure \ref{ej3}.

\subsection{Limit Circles, Limit Points and Spiral Branches}\label{subsec-limit}

\begin{figure}[ht]
\begin{center}
\centerline{$\begin{array}{cc}   \psfig{figure=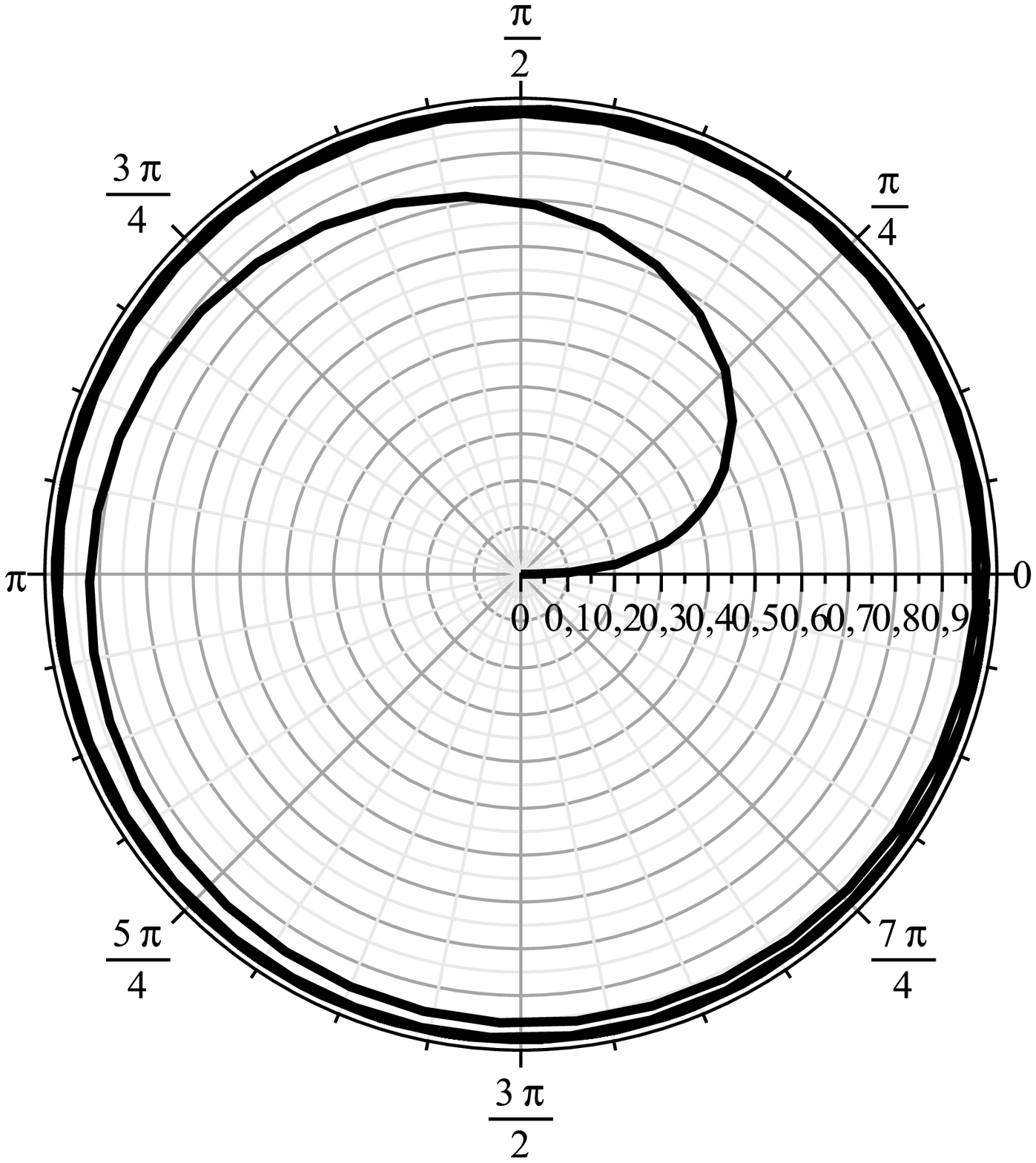,width=5cm,height=5cm} &
\psfig{figure=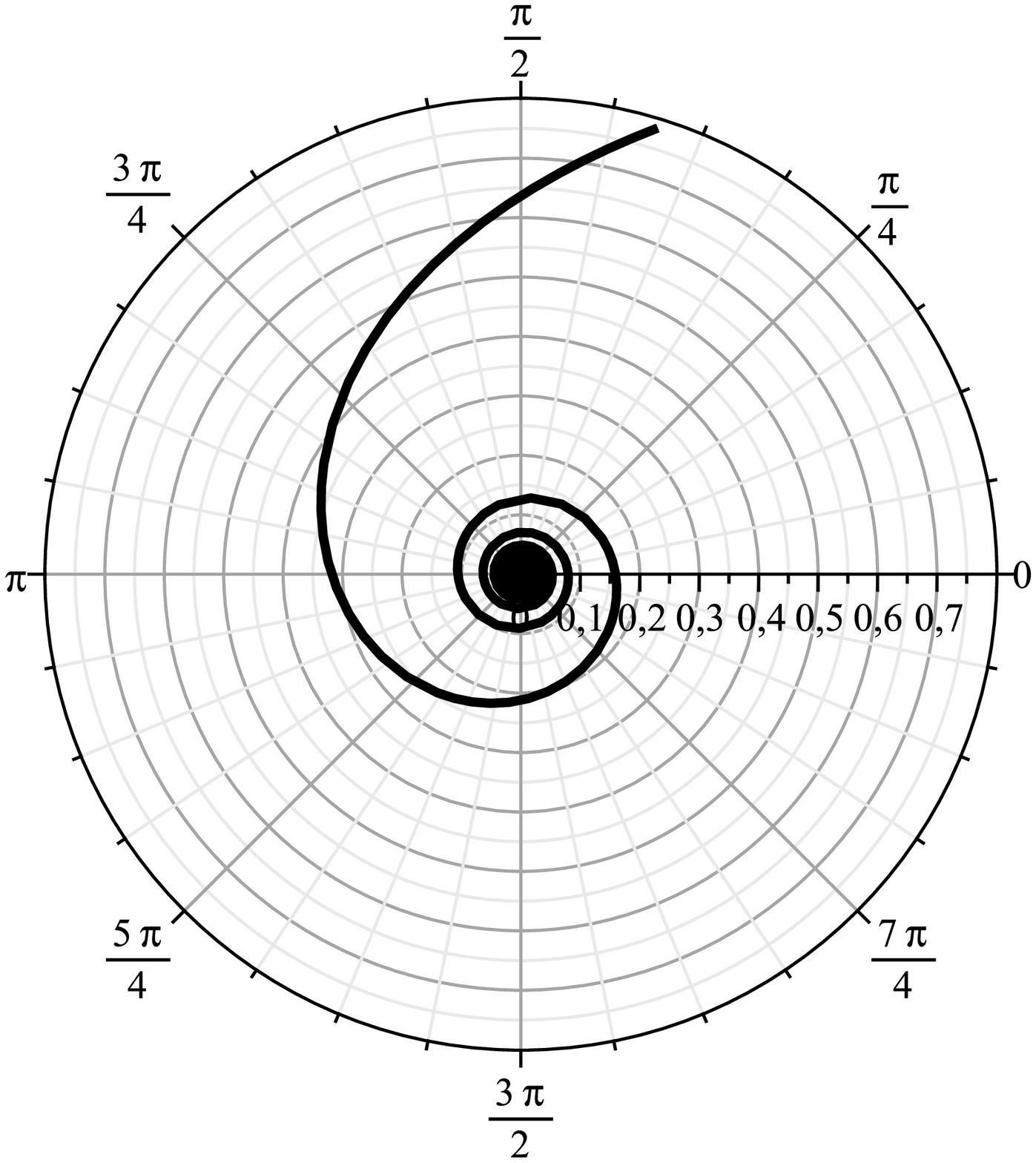,width=5cm,height=5cm}
\end{array}$}
\end{center}
\caption{Example of Limit Circle (left) and Limit Point  (right)}\label{clpl}
\end{figure}

Figure \ref{clpl} shows the plotting of the curves defined by $\varphi_1(t)=\left(\displaystyle{\frac{t^2}{t^2+1}, \frac{t^3}{t^2+1}}\right)$ for $t\in (0,6\pi)$ (left), and $\varphi_2(t)=(t,1/t)$ for $t\in(0,\pi/4)$ (right). In the first case, one sees that the curve winds infinitely around
the circle $r=1$, coming closer and closer to it. In the second case, the curve winds
infinitely around the origin of coordinates, somehow ``converging" to it. Finally,  Figure
\ref{ej2} shows a curve that winds infinitely around the origin, but getting further and further from it.  We will refer to the first situation by saying that here the curve exhibits a {\sf limit circle} ($r=1$
in this example).
In the second case, we will say that the origin is a {\sf limit point}; finally, we will refer to the third situation by saying
that the curve presents a {\sf spiral branch}. Along this subsection, we address these phenomena from a theoretical point of view and relate them to the appearance of infinitely many self-intersections.

\begin{definition} \label{def-limit-spiral}
Let $t_0\in {\Bbb R}\cup \{\pm \infty\}$. We say that ${\mathcal C}$ exhibits, for $t=t_0$:
\begin{enumerate}
\item  A {\sf limit circle} if $\mbox{lim}_{t\to t_0}r(t)=r_0\in {\Bbb R}$, $r_0\neq 0$, and $\mbox{lim}_{t\to t_0}\theta(t)=\pm \infty$.
\item  A {\sf limit point} at the origin if $\mbox{lim}_{t\to t_0}r(t)=0$, and $\mbox{lim}_{t\to t_0}\theta(t)=\pm \infty$.
\item   A {\sf spiral branch}t if $\mbox{lim}_{t\to t_0}r(t)=\pm \infty$, and $\mbox{lim}_{t\to t_0}\theta(t)=\pm \infty$
\end{enumerate}
In each case, we will say that $t_0$ {\it generates} the limit circle, the limit point or the spiral branch.
\end{definition}

Notice that limit points are degenerated cases of limit circles, namely when $r_0=0$. One might wonder if limit circles can be centered
at a point different from the origin, or if there can be limit points other than the origin. The answer, negative in both cases, is given by the
following theorem.

\begin{theorem} \label{no-limit}
 A curve rational in polar coordinates cannot have a limit circle centered at a point different from the origin, or a limit point at a point
different from the origin.
\end{theorem}

{\bf Proof.} If ${\mathcal C}$ had a limit circle centered at $P\neq (0,0)$, or a limit point $Q\neq (0,0)$, we would have infinitely
many local maxima and minima of $r(t)$. However, this cannot happen because $r(t)$ is a rational function,
and therefore the number of $t$-values fulfilling $r'(t)=0$ is finite. \qed


\begin{remark} \label{rem-2}  The same argument of Theorem \ref{no-limit} proves that
a curve rational in polar coordinates cannot have any other ``attractor"
different from the origin, or a circle centered at the origin.
\end{remark}

 Note that if ${\mathcal C}$ has infinitely many self-intersections, by Theorem \ref{th-3} the function $\theta(t)$ is not bounded and so there must be limit points, limit circles or spiral branches. More concretely, we have the following result.

\begin{proposition} \label{limit-self-int}
Let $I$ be a subset of ${\Bbb R}$ (not necessarily an interval) with infinitely many $t$-values generating self-intersections of
${\mathcal C}$. Then the closure of $I$ contains some $t$-value giving rise to either a limit point, or a limit circle, or a spiral branch of ${\mathcal C}$.
\end{proposition}


The converse of Proposition \ref{limit-self-int} is not true; for instance, in Example 2 we have a curve with a spiral branch for
$t_0= \infty$ without self-intersections. Now we want to characterize the situation when a limit point, a limit circle or a spiral branch have infinitely many close self-intersections. First we need the following definition.

\begin{definition} \label{close-inf}
We say that a limit circle, a limit point or a spiral branch generated by
$t_0\in {\Bbb R}\cup {\pm \infty}$ has {\sf infinitely many close self-intersections}, if there exists some real interval $I$ verifying:
\begin{enumerate}
\item $t_0 \in \overline{I}$;
\item $I$  does not contain any other $t$-value generating a limit circle, limit point or spiral;
\item  infinitely many $t\in I$ generate self-intersections of ${\mathcal C}$.
\end{enumerate}
\end{definition}

\begin{figure}[ht]
\begin{center}
\centerline{$\begin{array}{c}   \psfig{figure=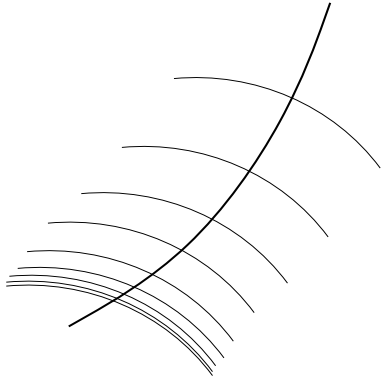,width=5cm,height=5cm}
\end{array}$}
\end{center}
\caption{Infinitely many close self-intersections}\label{imcs}
\end{figure}

Definition \ref{close-inf} is illustrated in Figure \ref{imcs}; the thin lines correspond to the branch generated by the interval $I$ appearing in Definition \ref{close-inf}.

\begin{theorem} \label{th-self-assoc}
Let $t_0\in {\Bbb R}\cup {\pm \infty}$ generating a limit circle, a limit point or a spiral branch. Then
\begin{enumerate}
\item If $t_0\in {\Bbb R}$, then $t_0$ has infinitely many close self-intersections if and only if $t=t_0$ is a horizontal asymptote of some of the curves $\xi_1(t,k)=0$ or $\xi_2(t,k)=0$.
\item  If $t_0=\pm \infty$, then it has infinitely many close self-intersections if and only if some of the curves $\xi_1(t,k)=0$ or $\xi_2(t,k)=0$ exhibits an infinite branch as $t\to \pm \infty$ which is not  a horizontal
asymptote (i.e. there exists a sequence of real points $(t_n,k_n)$ of the curve with $t_n\to \pm \infty$ and $k_n\to \pm \infty$).
\end{enumerate}
\end{theorem}

{\bf Proof.} We prove (1); the proof of (2) is similar. Assume that there exists an interval $I\subset {\Bbb R}$
satisfying Definition \ref{close-inf}.
Then the set of points of either $\xi_1(t,k)=0$ or $\xi_2(t,k)=0$ with $t\in I$ must be non-bounded in $k$.
This can only happen if there exists $t_a\in I$ such that $t=t_a$ is an asymptote of either $\xi_1(t,k)=0$ or $\xi_2(t,k)=0$. However, if $t=t_a$
is an asymptote then for any interval $I_a$ containing $t_a$, and not containing any other $t$-value where $\theta(t)$ is infinite, we can find
a non-bounded portion of either $\xi_1(t,k)=0$ or $\xi_2(t,k)=0$, therefore giving rise to infinitely many self-intersections of ${\mathcal C}$,
with $t$-values in $I_a$. So, from Proposition \ref{limit-self-int} we have that $t_0=t_a$. Using this same argument we can prove the converse
statement, i.e. if $t=t_0$ is an asymptote then any interval containing it has the desired properties. \qed

The detection of asymptotes of an algebraic curve is addressed for example in \cite{Zeng}.

\begin{corollary} \label{corol-self-assoc}
If $t=t_0$ has infinitely many close self-intersections, then any interval containing $t_0$ generates infinitely many self-intersections
of ${\mathcal C}$.
\end{corollary}

The above results are illustrated in the following examples.

{\bf Example 1 (cont.):} Recall that $\xi_1(t,k)=2k\pi$ and $\xi_2(t,k)=2t-2k\pi-\pi$. The second one has an infinite
branch  as $t\to \pm \infty$; so, from Theorem \ref{th-self-assoc} we deduce that every $t$-interval of the form $(-\infty,a)$ or $(b,\infty)$
contains infinitely many $t$-values generating self-intersections of ${\mathcal C}$.

\begin{figure}[ht]
\begin{center}
\centerline{$\begin{array}{ccc}   \begin{array}{c}\psfig{figure=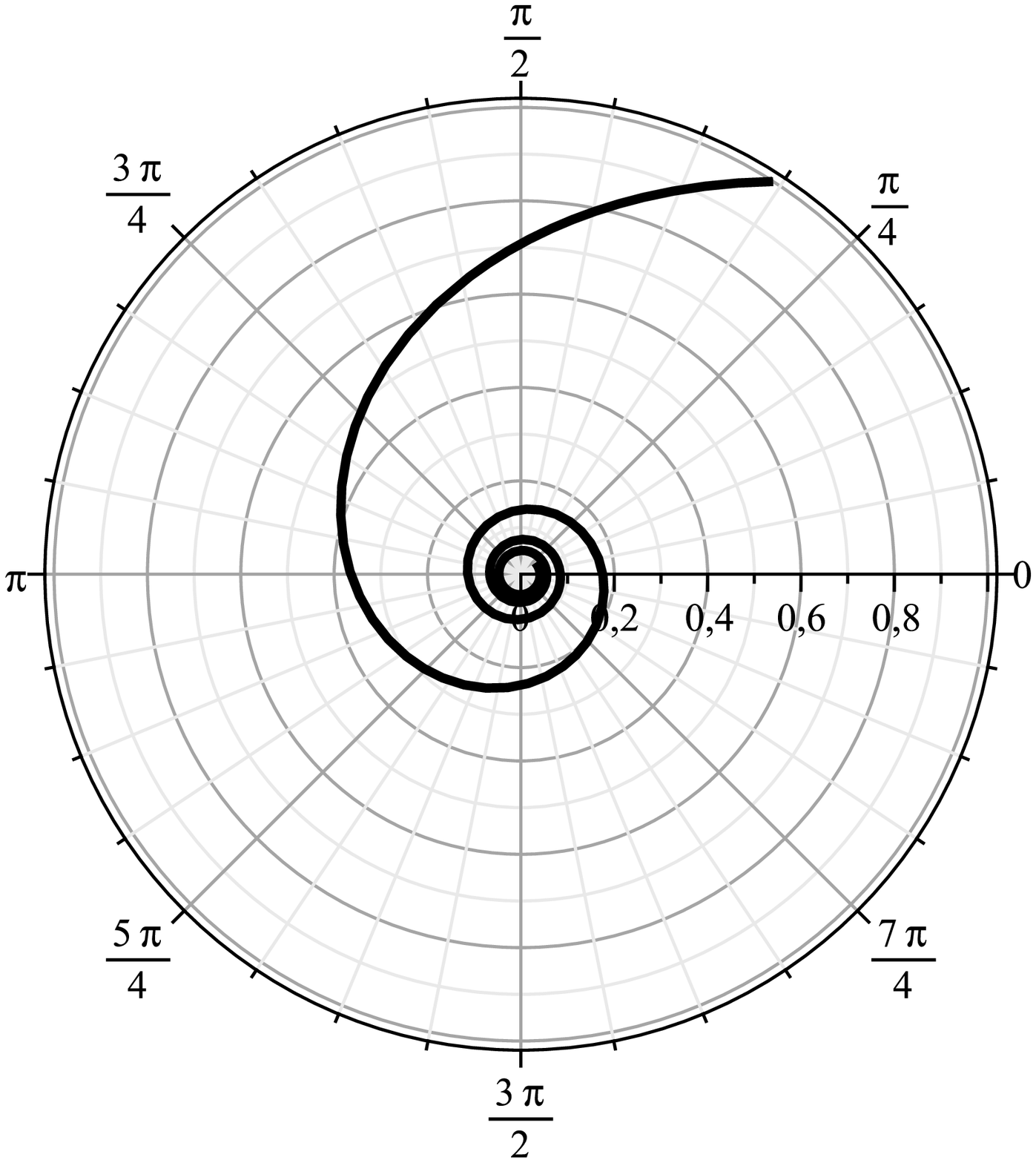,width=5cm,height=5cm} \\ t \in (1,5)\end{array}   &
 \begin{array}{c} \psfig{figure=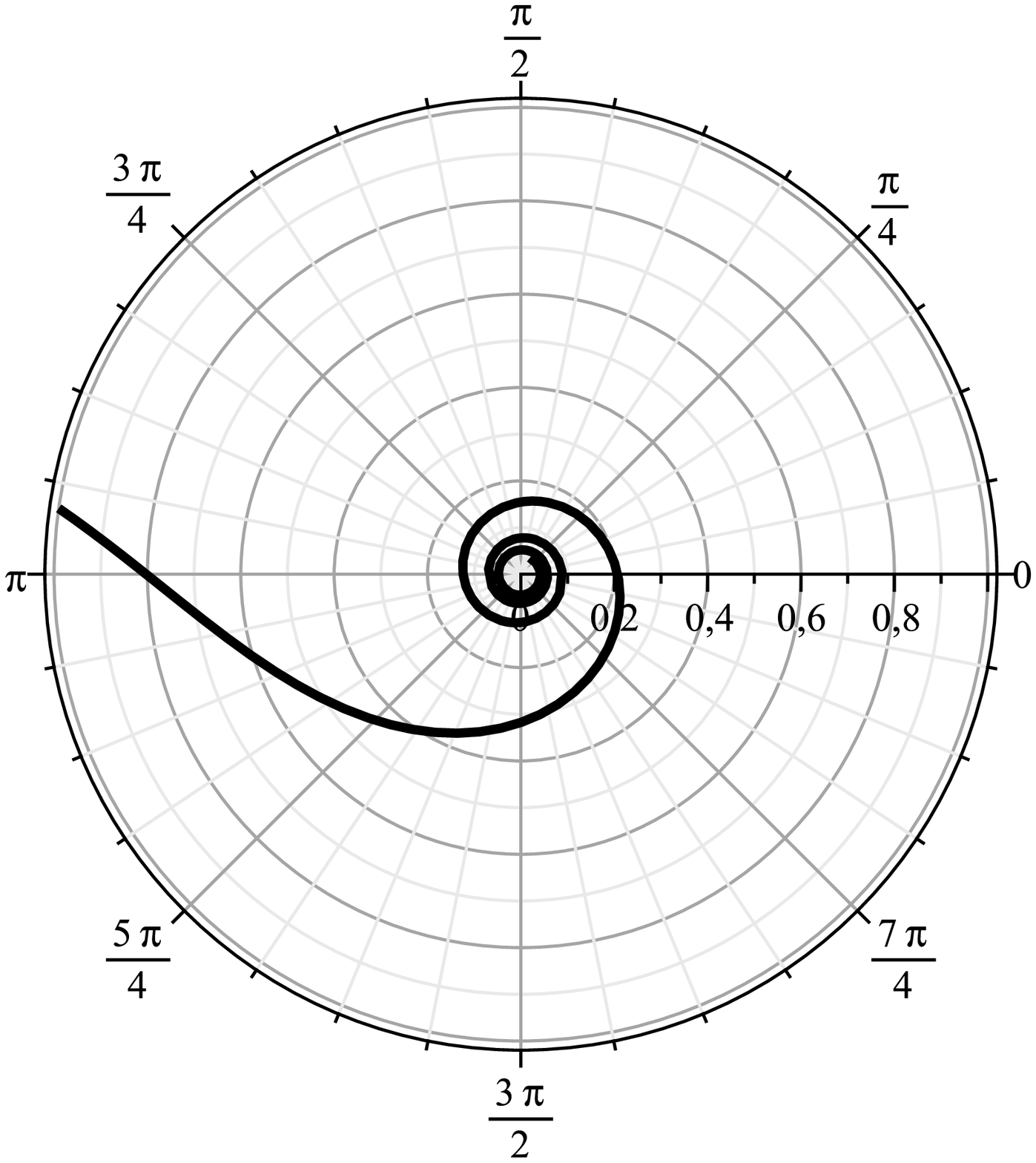,width=5cm,height=5cm}\\ t \in (-5,-1) \end{array} & \begin{array}{c}\psfig{figure=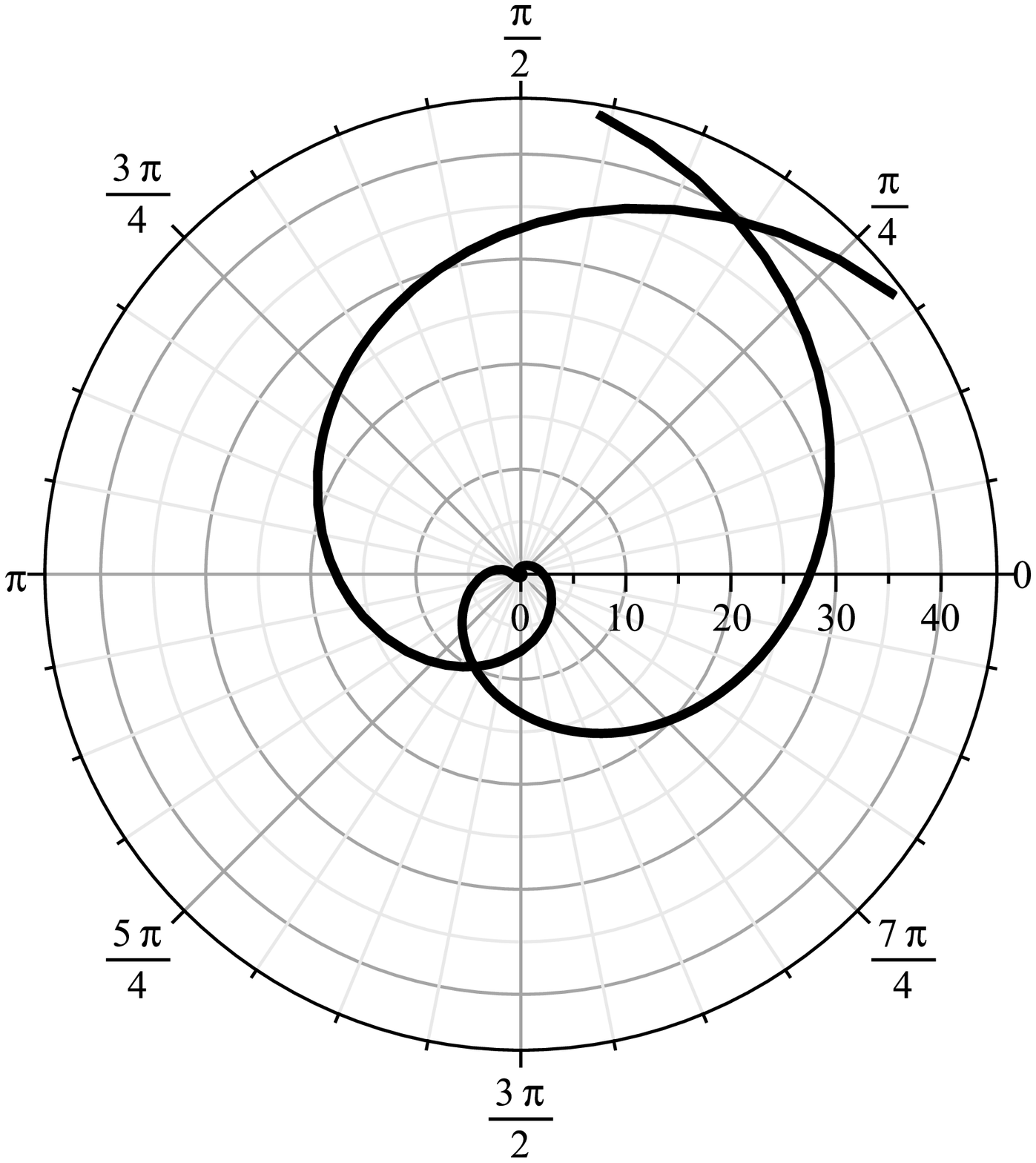,width=5cm,height=5cm} \\  t \in (-5,-0.15) \cup (0.15,5) \end{array}
\end{array}$}
\end{center}
\caption{$\varphi(t)=\displaystyle{\left(\frac{1}{t^2},\frac{t^3+t-1}{t}\right)}$ }\label{la7}
\end{figure}

{\bf Example 4.} Consider the curve defined by $\varphi(t)=\displaystyle{\left(\frac{1}{t^2},\frac{t^3+t-1}{t}\right)}$. This
curve has a spiral branch for $t=0$ and a limit point for $t\to \pm \infty$. Observe
Figure \ref{la7} where the curve is plotted for different values of $t$. A direct computation yields $\xi_1(t,k)=k(k\pi t+1)$ and
$$
\xi_2(t,k)= 4\,{t}^{6}-4\,\pi \, \left( 2\,k+1 \right) {t}^{4}-4\,{t}^{3}+{\pi }
^{2} \left( 2\,k+1 \right) ^{2}{t}^{2}+2\,\pi \, \left( 2\,k+1
 \right) t+2.
 $$

The curve $\xi_2(t,k)=0$ is
empty over the reals. However a factor of $\xi_1(t,k)=0$ corresponds to a hyperbola whose asymptotes are $t=0$ and $k=0$. Hence, from
statement (1) of Theorem \ref{th-self-assoc} we deduce that the spiral branch
generated by $t_0=0$ has infinitely many self-intersections.

\subsection{Asymptotes}\label{as}

It is classical (see for example \cite{DD}) that asymptotes in polar form correspond to values $t=t_0\in {\Bbb R}\cup \{\pm \infty\}$ such that:
\begin{itemize}
\item $\mbox{lim}_{t\to t_0}r(t)=\pm \infty$.
\item $\mbox{lim}_{t\to t_0}\theta(t)=\alpha\in {\Bbb R}$.
\item $\mbox{lim}_{t\to t_0}r(t)\cdot (\theta(t)-\alpha)\in {\Bbb R}$.
\end{itemize}
If the above conditions hold, then the asymptote is the line parallel to
the line with slope $\alpha$, at distance $\delta=\mbox{lim}_{t\to t_0}r(t)\cdot (\theta(t)-\alpha)$ of the origin.

\begin{figure}[ht]
\begin{center}
\centerline{ \psfig{figure=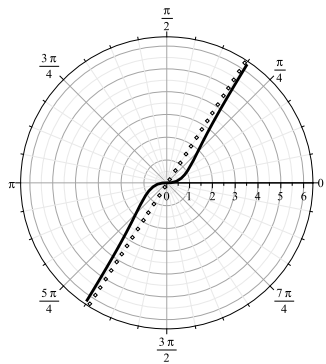,width=5cm,height=5cm} }
\end{center}
\caption{An asymptote}\label{f8}
\end{figure}

{\bf Example 5.} Consider the curve $\varphi_3(t)=\left(\displaystyle{t,\frac{t^2}{t^2+1}}\right)$. Then we have that $\lim_{t\to \pm
\infty}r(t)=\pm \infty$, $\lim_{t\to \pm \infty}\theta(t)=1$ and $\lim_{t\to \pm \infty}r(t)\cdot (\theta(t)-1)=0$. So, the line passing through
the origin and with slope 1, i.e. $y=x$, is an asymptote of the curve for $t\to \pm \infty$ (see Figure \ref{f8}).

\section{Visualizing Curves that are Rational in Polar Coordinates}

The goal of this section is to introduce an algorithm,  {\tt polares},  for visualizing the ``interesting" part of  ${\mathcal C}$, providing as well relevant information about the curve. We have implemented it in Maple 15 and tested it over several examples, some of which are presented in this section.

Next we describe the algorithm in the following different cases:
\begin{enumerate}
  \item $r(t)$ and $\theta(t)$ are both bounded
  \item $\theta(t)$ is bounded and $r(t)$ is not.
  \item  $r(t)$ is bounded and $\theta(t)$ is not.
  \item  $r(t)$ and $\theta(t)$ are both unbounded.
\end{enumerate}

Obviously, the first step of the algorithm is to detect the case we are in; then the algorithm proceeds accordingly.

\subsection{Algorithm {\tt polares} when $r(t)$ and $\theta(t)$ bounded}

{\tt Input:} A proper polar parametrization $\varphi(t)=(r(t),\theta(t))$. \newline
{\tt Output:}
 \begin{enumerate}
  \item Information about the existence of $P_{\infty}$.
  \item Information about the self-intersections.
  \item Plot  of $\varphi(t)$ for $t$ in ${\Bbb R}$ using the Maple command \texttt{polarplot}.
\end{enumerate}

Examples:

\begin{itemize}
  \item $\varphi_1(t)=( \frac{t}{1+t^2},\frac{t^2}{1+t^2})$
\end{itemize}

\texttt{\textcolor{red}{$>$polares([$\mathrm{\frac{t}{1+t^2},\frac{t^2}{1+t^2}}$]);}}\newline
\texttt{\textcolor{blue}{
 r and theta both bounded \newline
 Real point at the infinity such that (r, theta)=[0, 1] and the point is [0, 0] \newline
The point at infinity (0,0) is reached 1 times in R, so self-intersection at the origen}}

\begin{figure}[ht]
\begin{center}
\centerline{ \psfig{figure=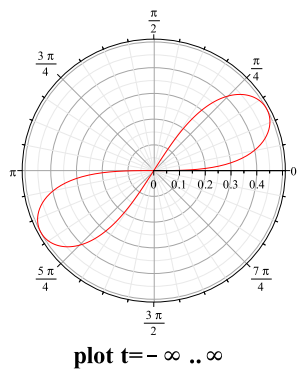,width=5cm,height=5cm} }
\end{center}
\caption{$\varphi_1(t)$}\end{figure}
\begin{itemize}
  \item $\varphi_2(t)=( \frac{t}{1+t^2},\frac{t^2+14}{1+t^2})$
  \end{itemize}

  \texttt{\textcolor{red}{$>$polares([$\mathrm{\frac{t}{1+t^2},\frac{t^2+14}{1+t^2}}$]);}}\newline
\texttt{\textcolor{blue}{  r and theta both bounded \newline
Real point at the infinity such that (r, theta)=[0, 1] and the point is [0, 0] \newline
The point at infinity (0,0) is reached 1 times in R, so self-intersection at the origen \newline
System (1) gives self-intersections for  k in [[ -2,2]], k$<>$0\newline
System (2) gives self-intersections for k in [[ -2,1]]}}

  \begin{figure}[ht]
\begin{center}
\centerline{ \psfig{figure=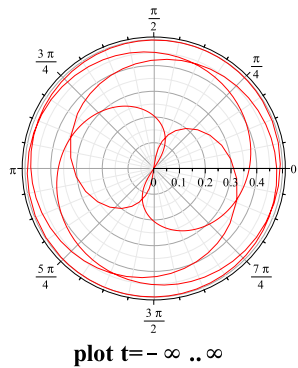,width=5cm,height=5cm} }
\end{center}
\caption{$\varphi_2(t)$}
\end{figure}

\subsection{Algorithm {\tt polares} when $\theta(t)$ is bounded and $r(t)$ not}\label{3.2}

{\tt Input:} A proper polar parametrization $\varphi(t)=(r(t),\theta(t))$. \newline
{\tt Output:}
 \begin{enumerate}
  \item Information about the existence of $P_{\infty}$.
  \item Information about the self-intersections.
  \item Information about the existence of asymptotes.
  \item \label{plot}Plot of $\varphi(t)$ using the Maple command \texttt{polarplot}:
  \begin{enumerate}
  \item If there are not values of $t$ generating asymptotes, then we plot the curve for $t \in  (-\infty,\infty)$;

  \item Otherwise, let $T_1$ be the set of values of $t$ generating asymptotes, let $T_2$ be the set of real values of $t$ such that $r(t)=0$ and, if $|\theta(t)|<2\,\pi$ for $t \in  {\Bbb R}$, let $T_3$ be the real values of $t$ generating the maximum and minimum of $\theta(t)$.
\newline
Let $P:=T_1 \cup T_2   \cup T_3 = \{t_1, \ldots,t_m\}  $  with $t_1<\ldots<t_m$.
  \newline
Now,

\begin{enumerate}
  \item If $\{t_1,t_m\}\neq \{\infty,-\infty\}$, then $P_\infty \in  {\Bbb R}^2$ and its plot may be of interest. So we add
$ t_0:= -\infty$  and $t_{m+1}:=\infty$ to the set $P$. 
 \newline 
For every $t_i \in P,\, 1<i<m$, we plot the curve $\varphi(t)$  for   $t\in(\frac{t_i+t_{i-1}}{2},\frac{t_i+t_{i+1}}{2})$. For  $t$ in  $(\frac{t_i+t_{i-1}}{2},t_i)$, the color of the plot will be red and for $( t_i,\frac{t_i+t_{i+1}}{2})$ the color will be blue. 
  \newline
As for $t_1$ and  $t_m$, we plot $\varphi(t)$ in the ranges $( 2t_1- \frac{t_1+t_2}{2}, \frac{t_1+t_2}{2})$ and  $(\frac{t_m-t_{m-1}}{2}, 2t_m- \frac{t_m+t_{m+1}}{2})$.
  \newline
  As for $t_0$ and  $t_{m+1}$, we plot $\varphi(t)$ in the ranges $(   -\infty, 2t_1- \frac{t_1+t_2}{2})$ and  $(  2t_m- \frac{t_m+t_{m+1}}{2}, \infty)$.

\item  If $\{t_1,t_m\}=\{\infty,-\infty\}$, then for every $t_i \in P,\, 2<i<m-1$, we plot the curve $\varphi(t)$ in the range  $(\frac{t_i+t_{i-1}}{2},\frac{t_i+t_{i+1}}{2})$. For $t$ in  $(\frac{t_i+t_{i-1}}{2},t_i)$, the color of the plot will be red and for $( t_i,\frac{t_i+t_{i+1}}{2})$ the color will be blue.
\newline
As for $t_2$ and  $t_{m-1}$, we plot $\varphi(t)$  for the ranges $(t_2-10,\frac{t_2+t_3}{2})$ and  $(\frac{t_{m-1}-t_{m-2}}{2}, t_{m-1}+10)$.
\newline
Finally, as for $\{\infty,-\infty\}$, we plot $\varphi(t)$  for the ranges  $(t_2-20,t_2-10)$ and  $(  t_{m-1}+10,t_{m-1}+20)$.

\end{enumerate}

%

\end{enumerate}
Let us point out that we border the values of $t$ generating the asymtotes for plotting.

\end{enumerate}

Examples:

\begin{itemize}
  \item $ \varphi_3(t)=( \frac{t^2}{t^2-11\,t+30},\frac{t^2+78}{1+t^2})$
\end{itemize}

\texttt{\textcolor{red}{$>$polares([$\mathrm{t^2/(t^2-11\,t+30) , (t^2+78)/(t^2+1)}$]);}}\newline
\texttt{\textcolor{blue}{
r unbounded and theta bounded \newline
Real point at the infinity such that (r, theta)=[1, 1] and the point is [cos(1), sin(1)]\newline
Point at infinity is not reached with k=0\newline
Point at infinity is not reached with k$<>$0\newline
System (1) gives self-intersections for  k in [[ -2,2]], k$<>$0\newline
The values of t generating asymptotes are  $\{5., 6.\}$\newline
Values of t considered in the plot $\{-\infty,0., 5., 6.\infty\}$}}

 \begin{figure}[ht]
 $$
 \begin{array}{ccc}
$$
 { \psfig{figure=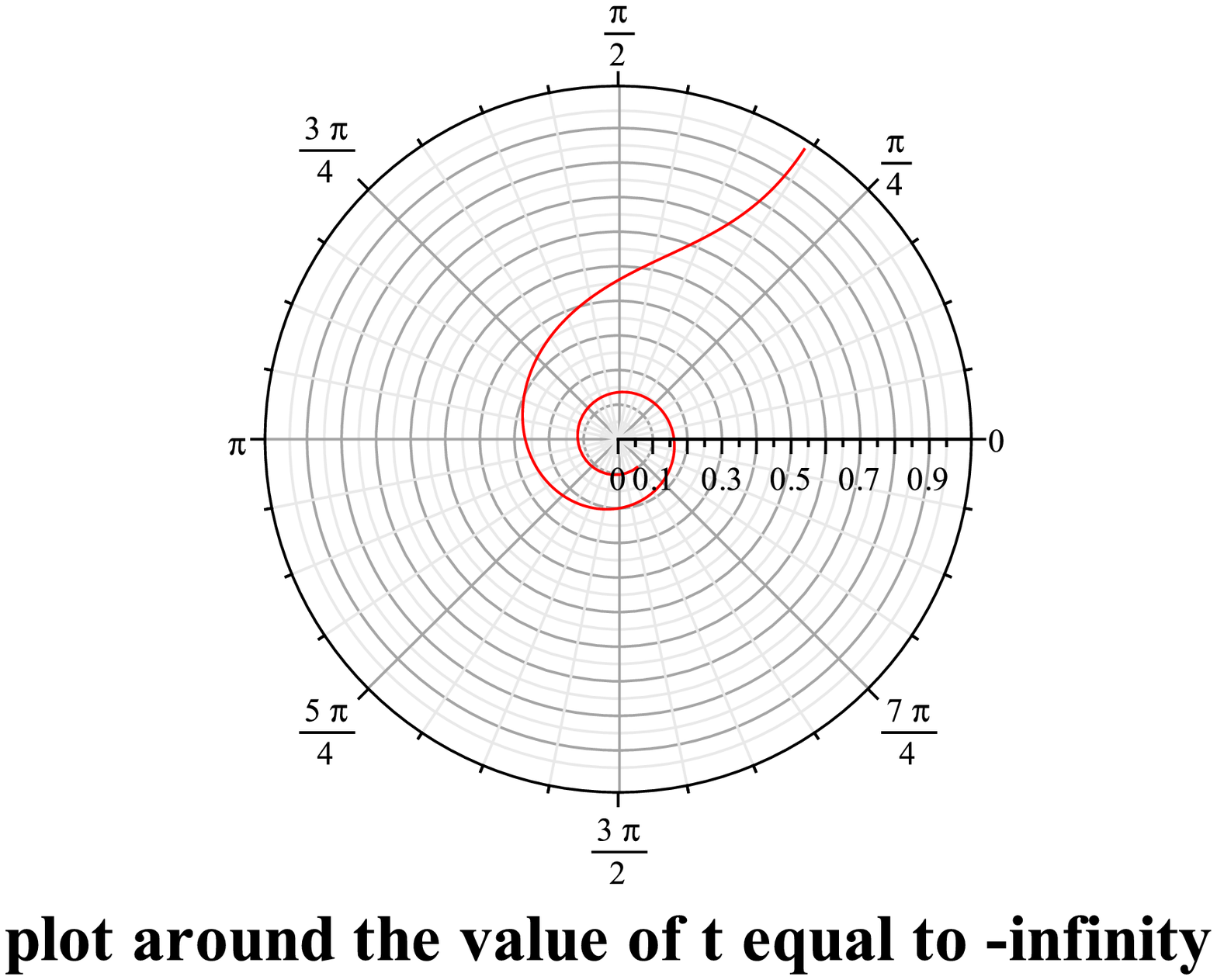,width=5cm,height=4cm} }&
 { \psfig{figure=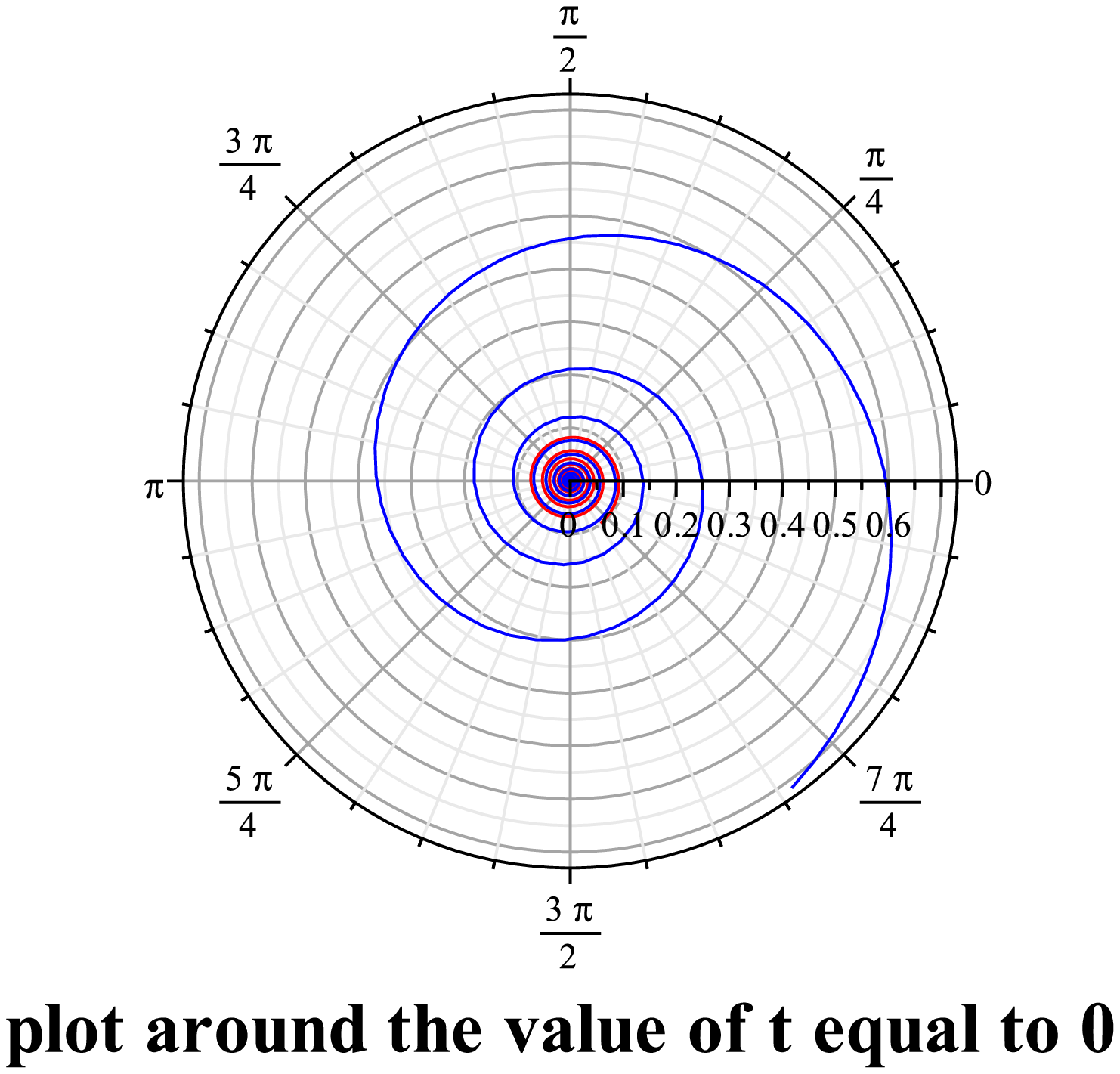,width=5cm,height=4cm} }&
 { \psfig{figure=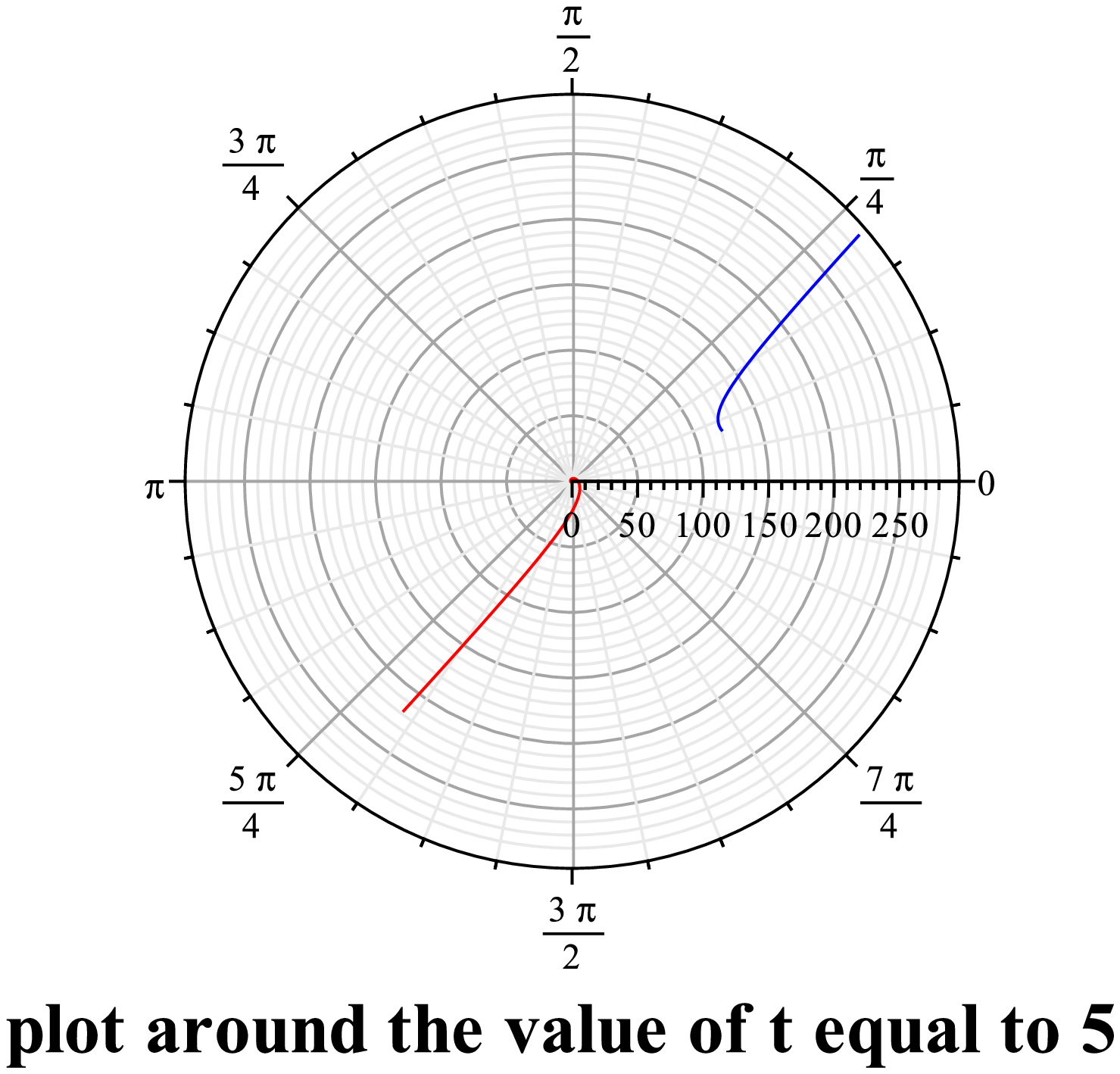,width=5cm,height=4cm} }\\
 { \psfig{figure=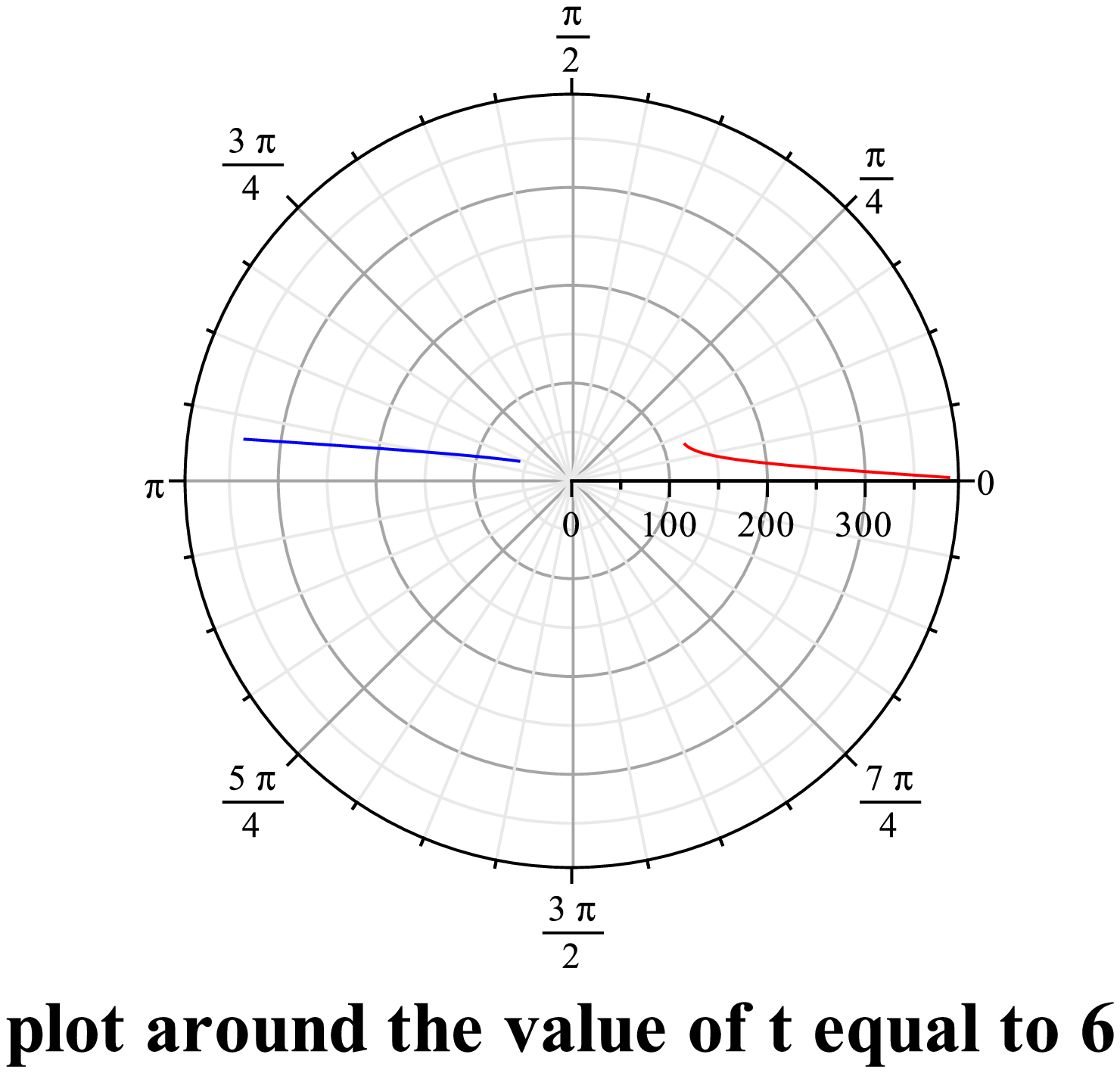,width=5cm,height=4cm} }&
 { \psfig{figure=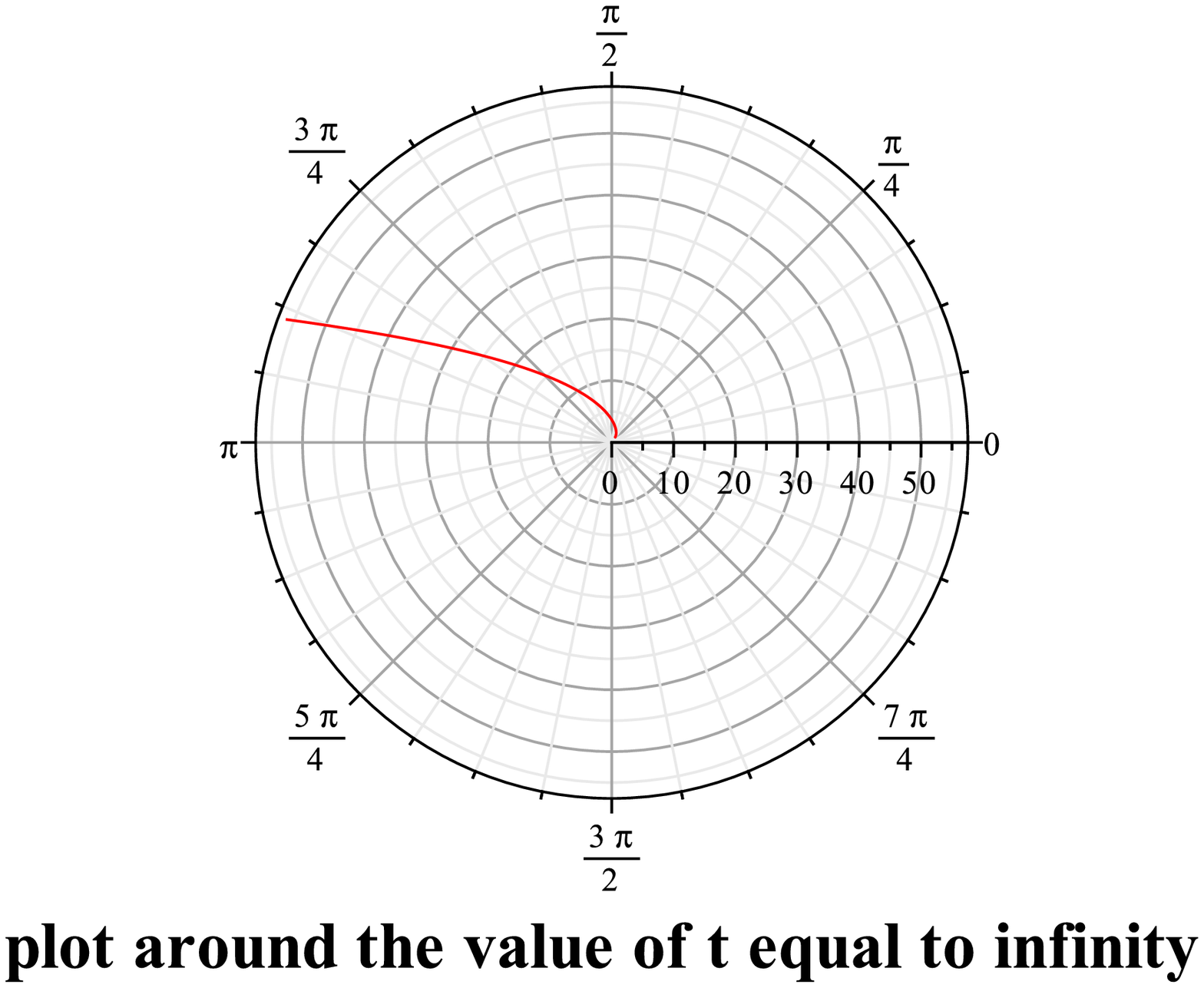,width=5cm,height=4cm} }&
  \end{array}
$$
\caption{ $\varphi_3(t)$}
\label{12}
\end{figure}

\begin{itemize}
  \item $ \varphi_4(t)=( t,\frac{t^2+14}{1+t^2})$
\end{itemize}

\texttt{\textcolor{red}{$>$polares([$\mathrm{t, (t^2+14)/(t^2+1)}$]);}}\newline
\texttt{\textcolor{blue}{
r unbounded and theta bounded \newline
There is no point at infinity \newline
Values of t generating asymptotes $[\infty, -\infty]$   \newline
Values of t considered in the plot $\{0., \infty, -\infty\}$
}}

   \begin{figure}[ht]
 $$\begin{array}{ccc}
 { \psfig{figure=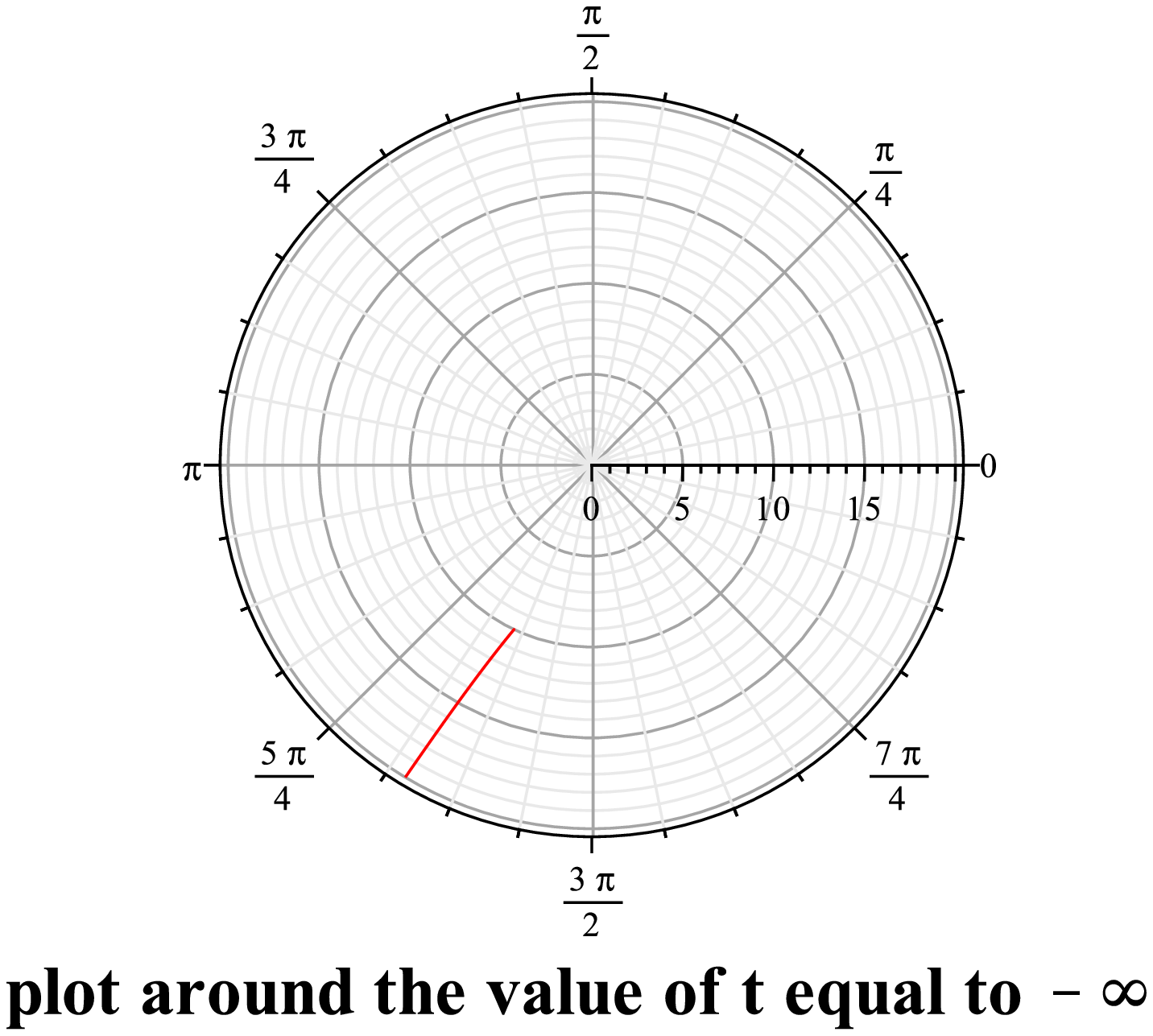,width=5.5cm,height=4cm} }&
 { \psfig{figure=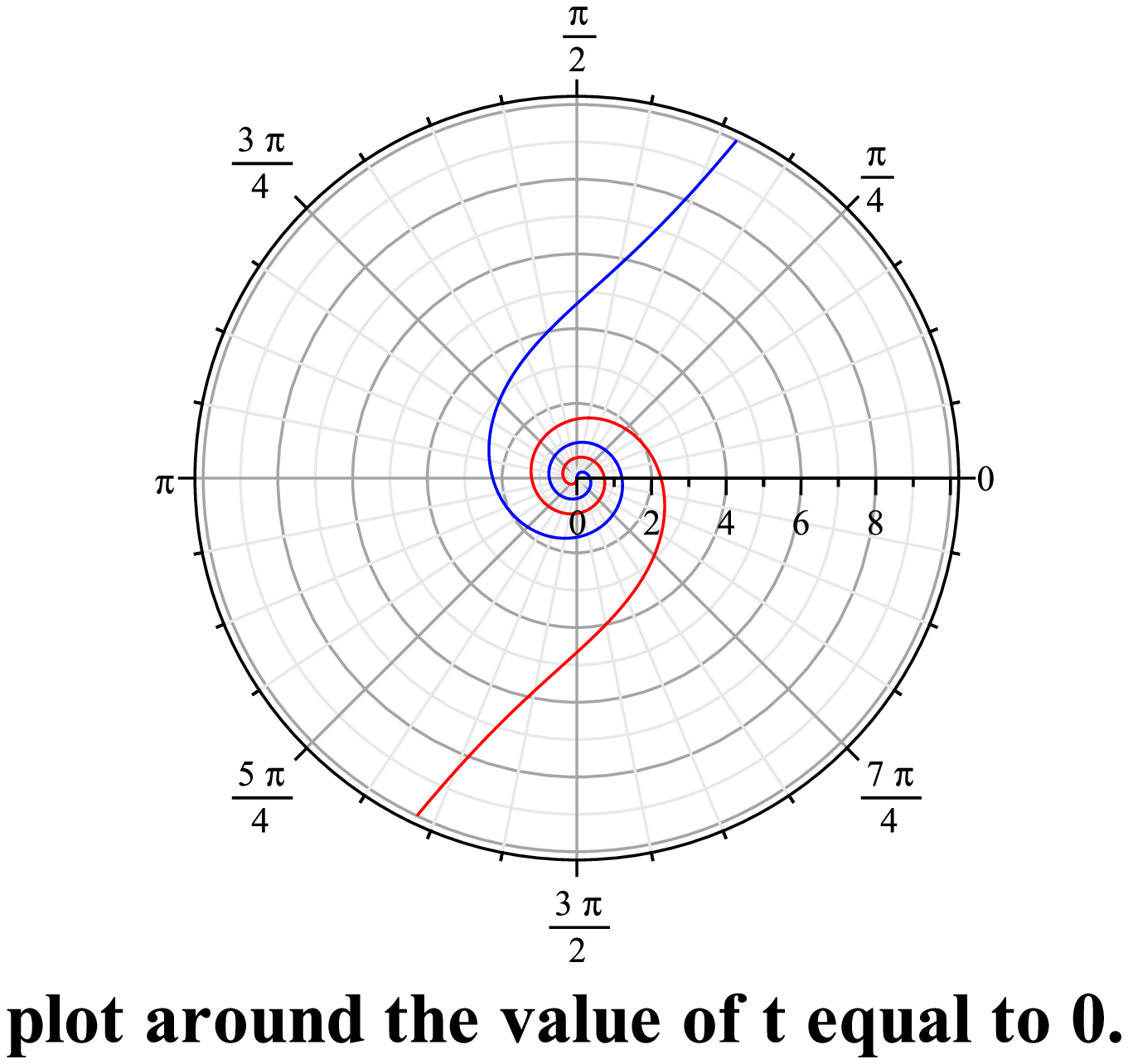,width=5.5cm,height=4cm} }&
 { \psfig{figure=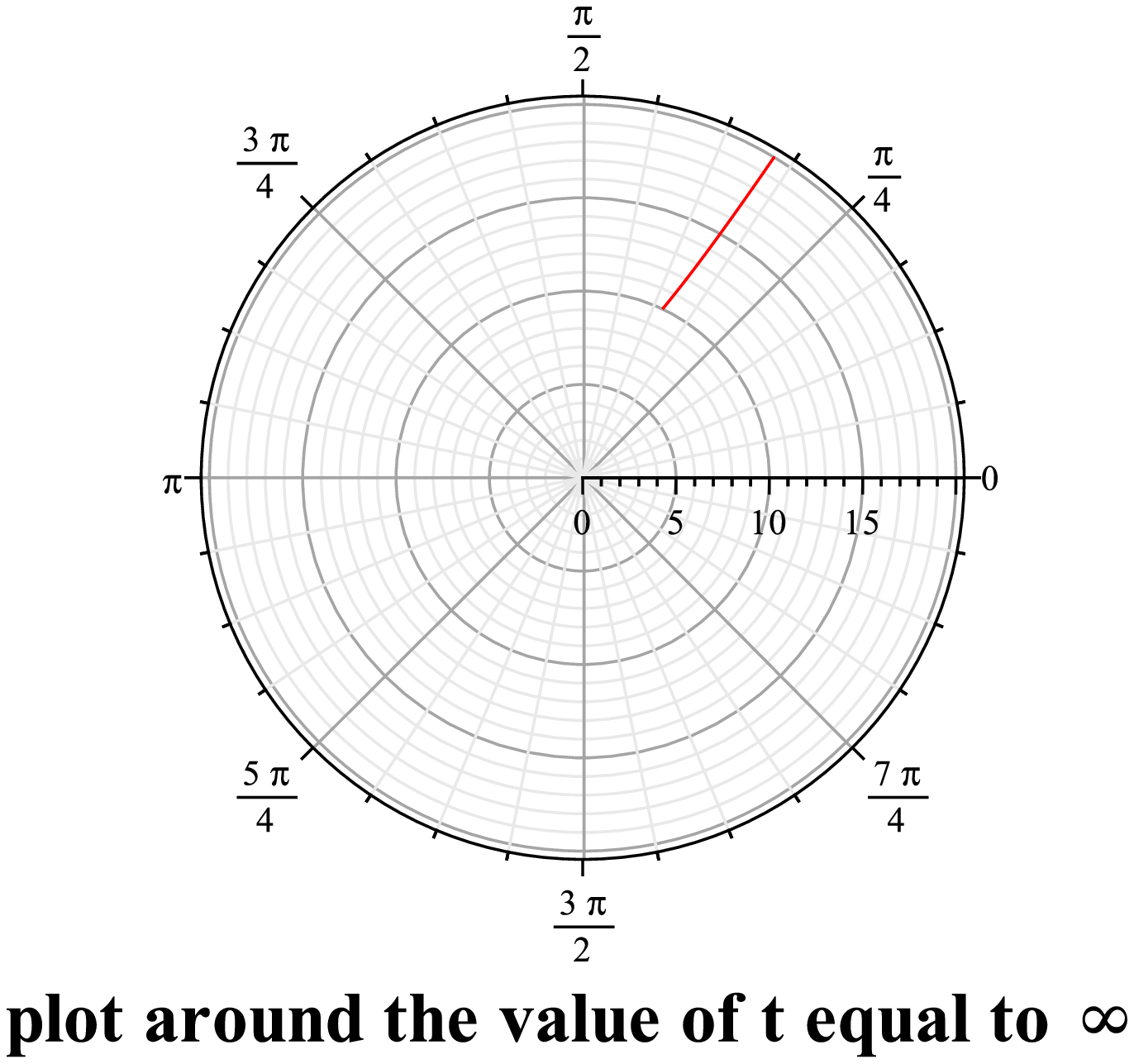,width=5.5cm,height=4cm} }
\end{array}
$$
\caption{ $ \varphi_4(t)$}
\label{13}
\end{figure}

\subsection{Algorithm {\tt polares} when $r(t)$ is bounded and $\theta(t)$ not}

 {\tt Input:} A proper polar parametrization $\varphi(t)=(r(t),\theta(t))$. \newline
{\tt Output:}
 \begin{enumerate}
  \item Information about the existence of $P_{\infty}$.
  \item Information about the existence of limit circles.
  \item Information about the existence of limit points.
  \item Information about the self-intersections.
  \item Plot  of $\varphi(t)$ for $t$ in ${\Bbb R}$ using the Maple command \texttt{polarplot}.

 \begin{enumerate}
  \item[] Let $T_1$ be the set of values of $t$ generating limit circles, let $T_2$ be the set of values of $t$ generating limit points,
  let $T_3$ be the set of real values of $t$ such that $r(t)=0$ and let $T_4$ be the real values of $t$ generating the maximum of $|r(t)|$.
\newline
Let $P:=T_1 \cup T_2 \cup T_3 \cup T_4  = \{t_1, \ldots,t_m\}$  with $t_1<\ldots<t_m$.
   Then we proceed similar to the case (\ref{plot}) of Section \ref{3.2}.  Now we border the values of $t$ generating limit circles for plotting.

  \end{enumerate}

\end{enumerate}

Example:

\begin{itemize}
  \item $\varphi_5(t)=(t^2/(t^2+1),t^3/(t^2+1) )$

\end{itemize}

\texttt{\textcolor{red}{$>$polares([$\mathrm{t^2/(t^2+1),t^3/(t^2+1)}$]);}}\newline
\texttt{\textcolor{blue}{r bounded and theta unbounded \newline
There is no point at infinity \newline
Values of t generating limit circles [$-\infty, \infty$] \newline
There are infinitely many self-intersections \newline 
t=infinity has infinitely many close self-intersections \newline
There are no limit points    \newline
Values of t considered in the plot $\{0., \infty, -\infty\}$
}}

 \begin{figure}[ht]
 $$
 \begin{array}{ccc}
 { \psfig{figure=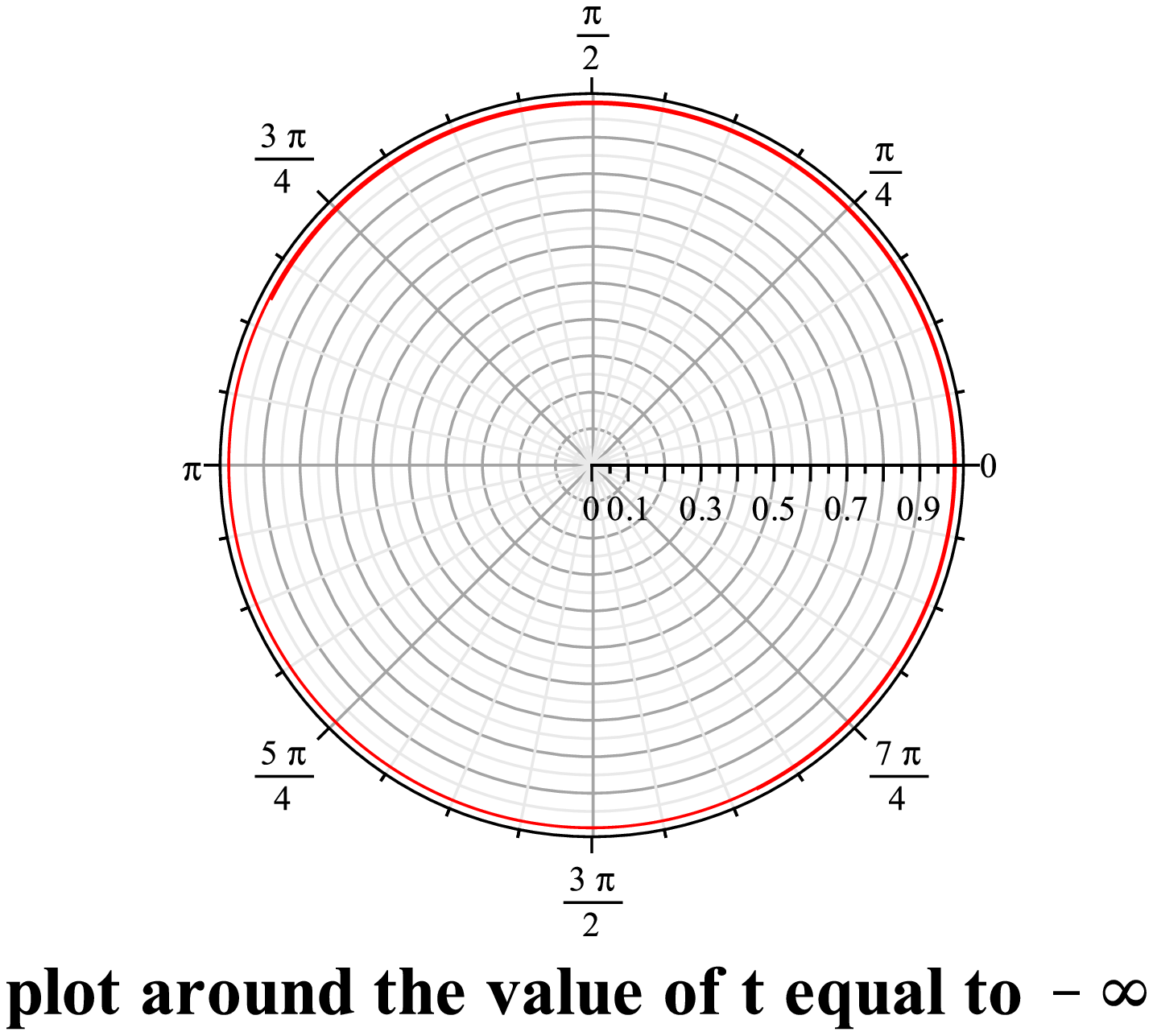,width=5.5cm,height=4cm} }&
 { \psfig{figure=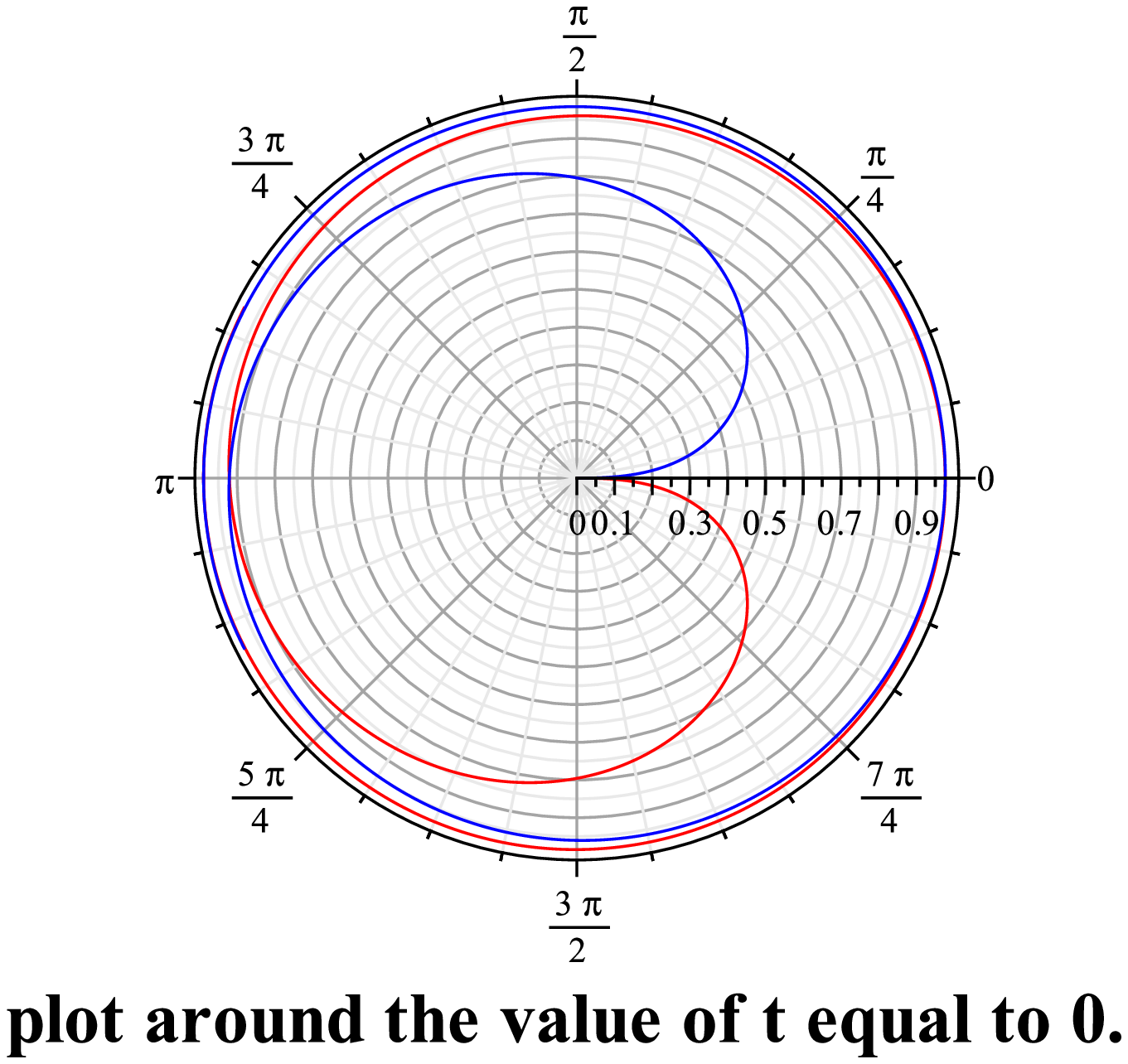,width=5.5cm,height=4cm} }&
 { \psfig{figure=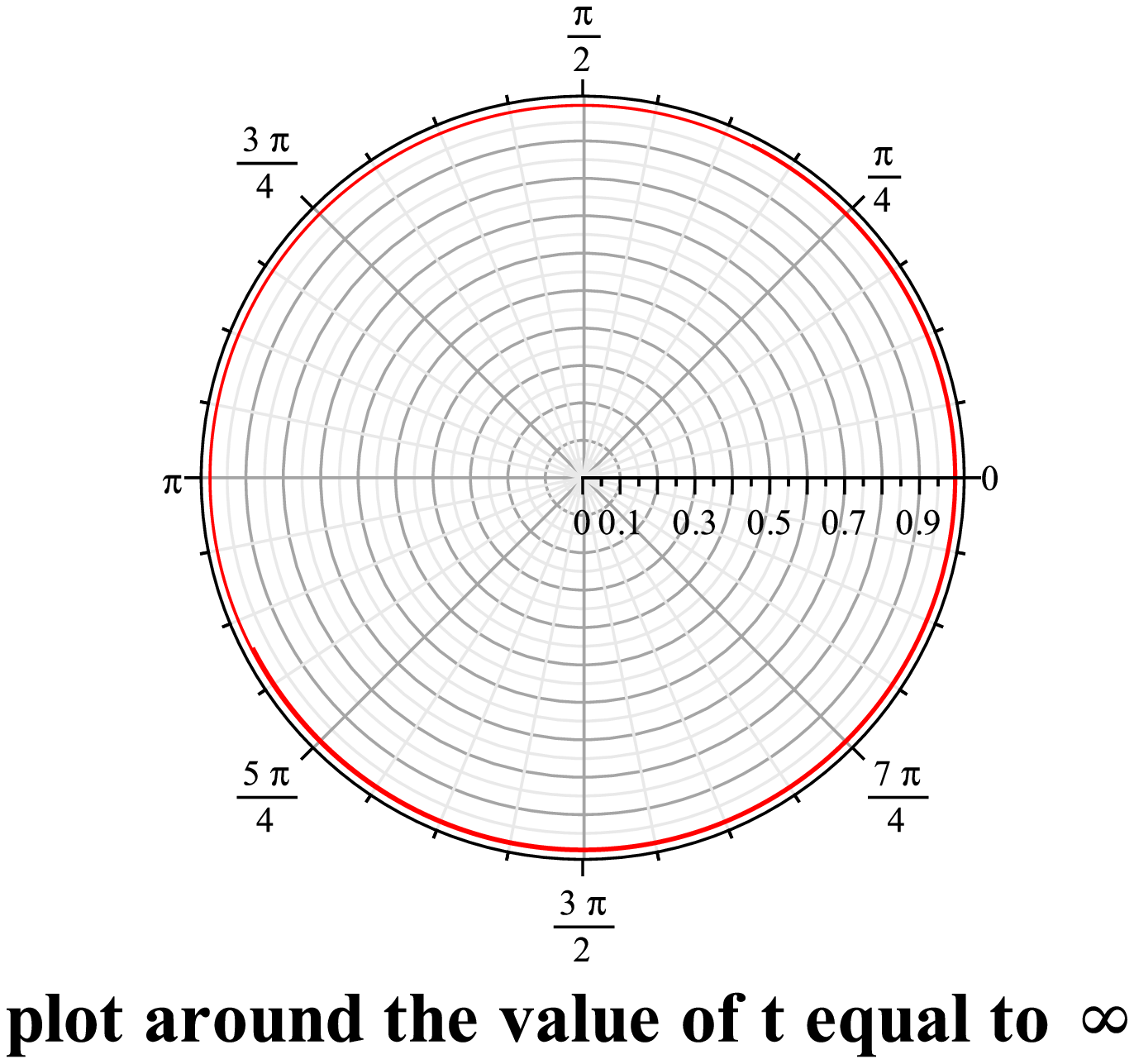,width=5.5cm,height=4cm} }
 \end{array}
 $$
\caption{ $\varphi_5(t)$}
\end{figure}

\subsection{Algorithm {\tt polares} when both $r(t)$ and $\theta(t)$ are unbounded}

 {\tt Input:} A proper polar parametrization $\varphi(t)=(r(t),\theta(t))$. \newline
{\tt Output:}
 \begin{enumerate}
  \item Information about the existence of $P_{\infty}$.
  \item Information about the existence of limit circles.
  \item Information about the existence of limit points.
  \item Information about the existence of spiral branches.
    \item Information about the existence of asymptotes.
  \item Information about the self-intersections.
  \item Plot  of $\varphi(t)$ for $t$ in ${\Bbb R}$ using the Maple command \texttt{polarplot}.

 \begin{enumerate}
  \item[] Let $T_1$ be the set of values of $t$ generating limit circles, let $T_2$ be the set of values of $t$ generating limit points,
  let $T_3$ be the set of values of $t$ generating spiral branches, let $T_4$ be the set of values of $t$ generating asymptotes and let $T_5$ be real the values of $t$ such that $r(t)=0$
\newline
Let  $T_1 \cup T_2 \cup T_3 \cup T_4 \cup T_5  = \{t_1, \ldots,t_m\}$  with $t_1<\ldots<t_m$.
\newline
Then we proceed similar to the case (\ref{plot}) of Section \ref{3.2}. Here we border the values of $t$ generating asymptotes, limit circles and spiral branches for plotting.
  \end{enumerate}

\end{enumerate}

Example:
\begin{itemize}
  \item $ \varphi_6(t)=(t,(t^3+1)/({t}^{2}-3\,t+2) )$
\end{itemize}

\texttt{\textcolor{red}{$>$polares([$\mathrm{t,(t^3+1)/({t}^{2}-3\,t+2)}$]);}}\newline
\texttt{\textcolor{blue}{r  and theta both unbounded \newline
There is no point at infinity \newline
Values of t generating limit circles  [1., 2.]  \newline
There are no limit points    \newline
Values of t generating spiral branches [$-\infty, \infty$] \newline
There are not values of t generating asymptotes \newline
There are infinitely many self-intersections \newline 
t=1 has infinitely many close self-intersections \newline
t=2 has infinitely many close self-intersections \newline
t=infinity has infinitely many close self-intersections \newline
Values of t considered in the plot $\{-\infty, 0.,1., 2., \infty \}$
}}

 \begin{figure}[ht]
 $$
 \begin{array}{ccc}
$$
 { \psfig{figure=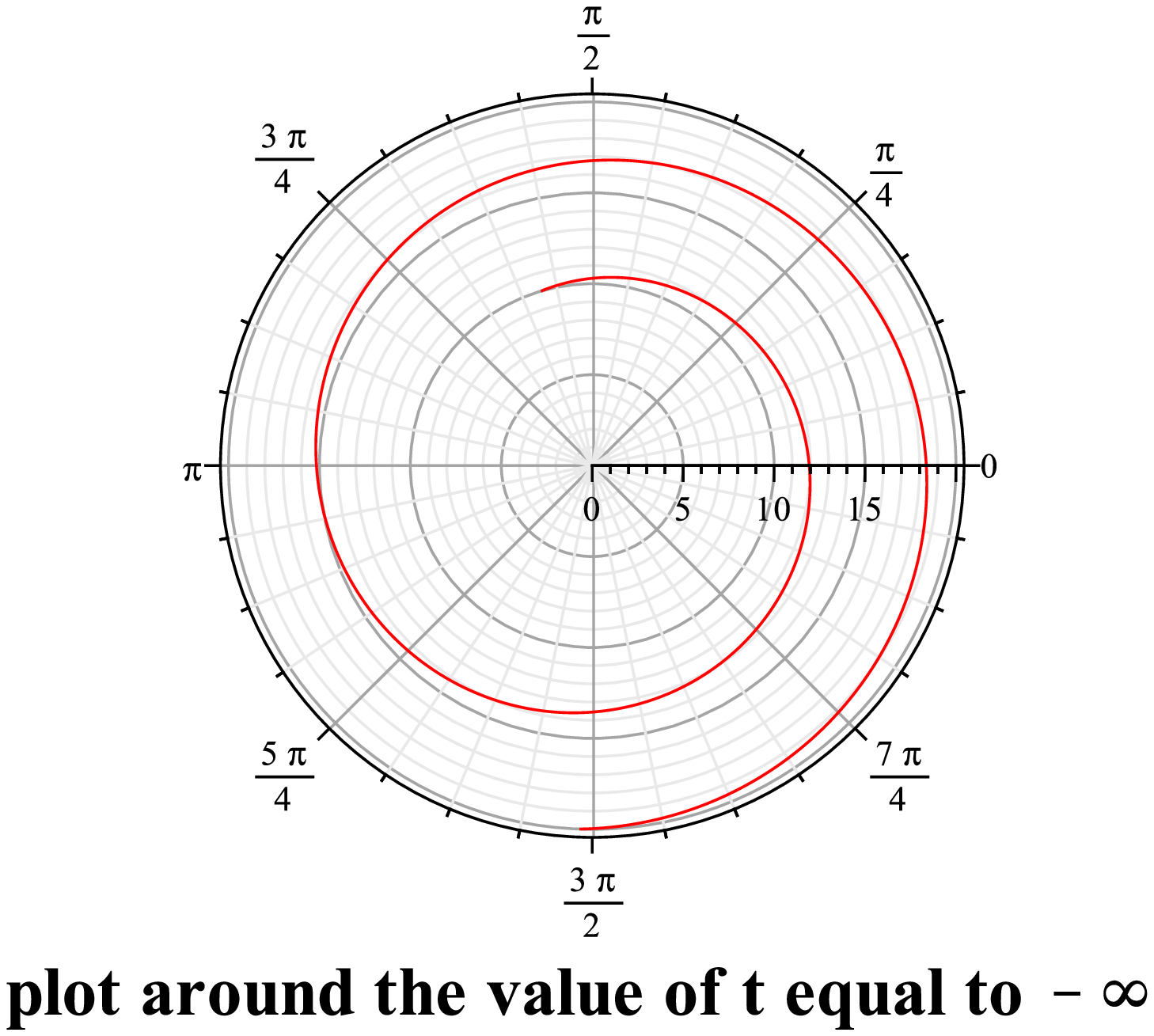,width=5.5cm,height=4cm} }&
 { \psfig{figure=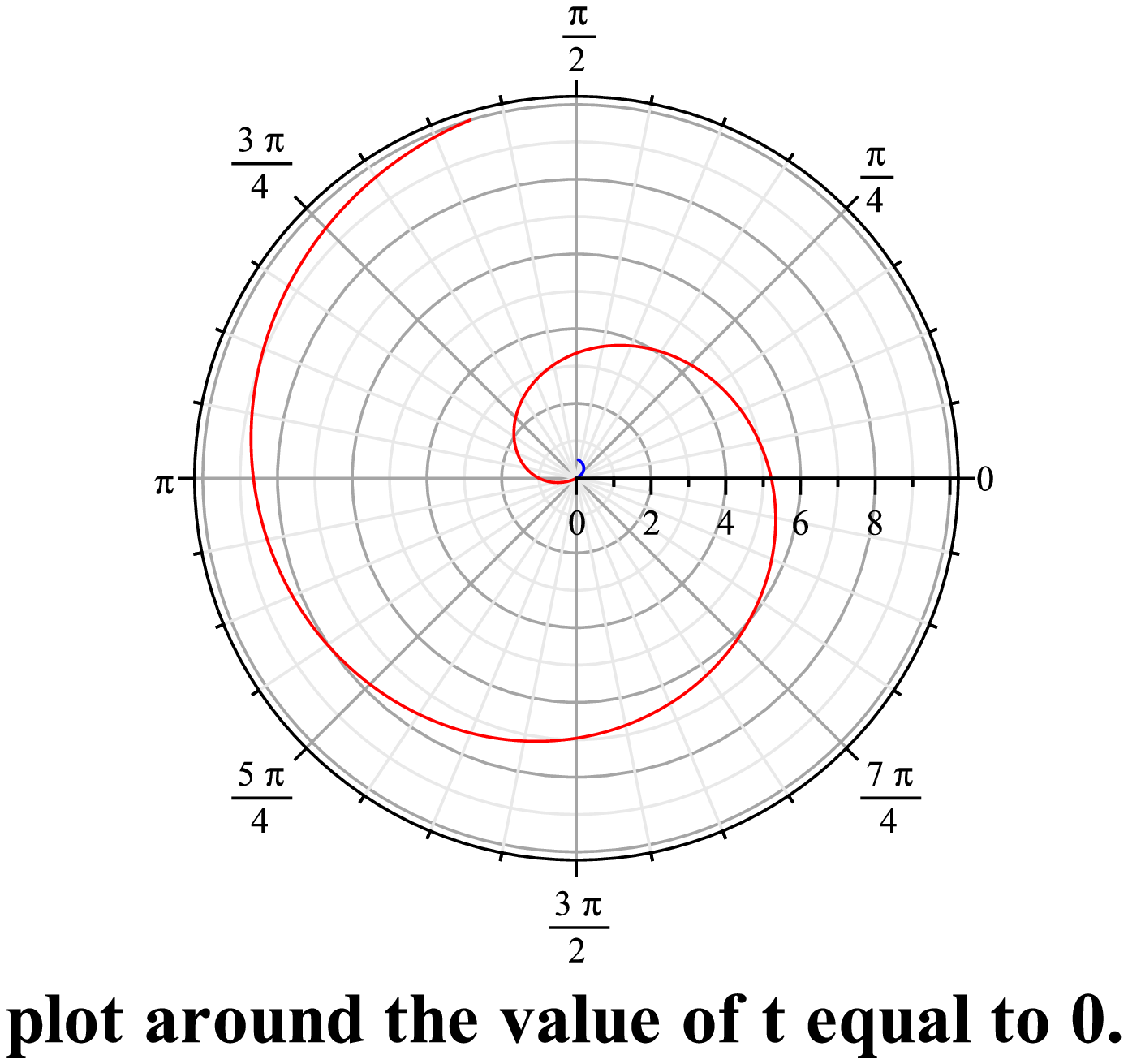,width=5.5cm,height=4cm} }&
 { \psfig{figure=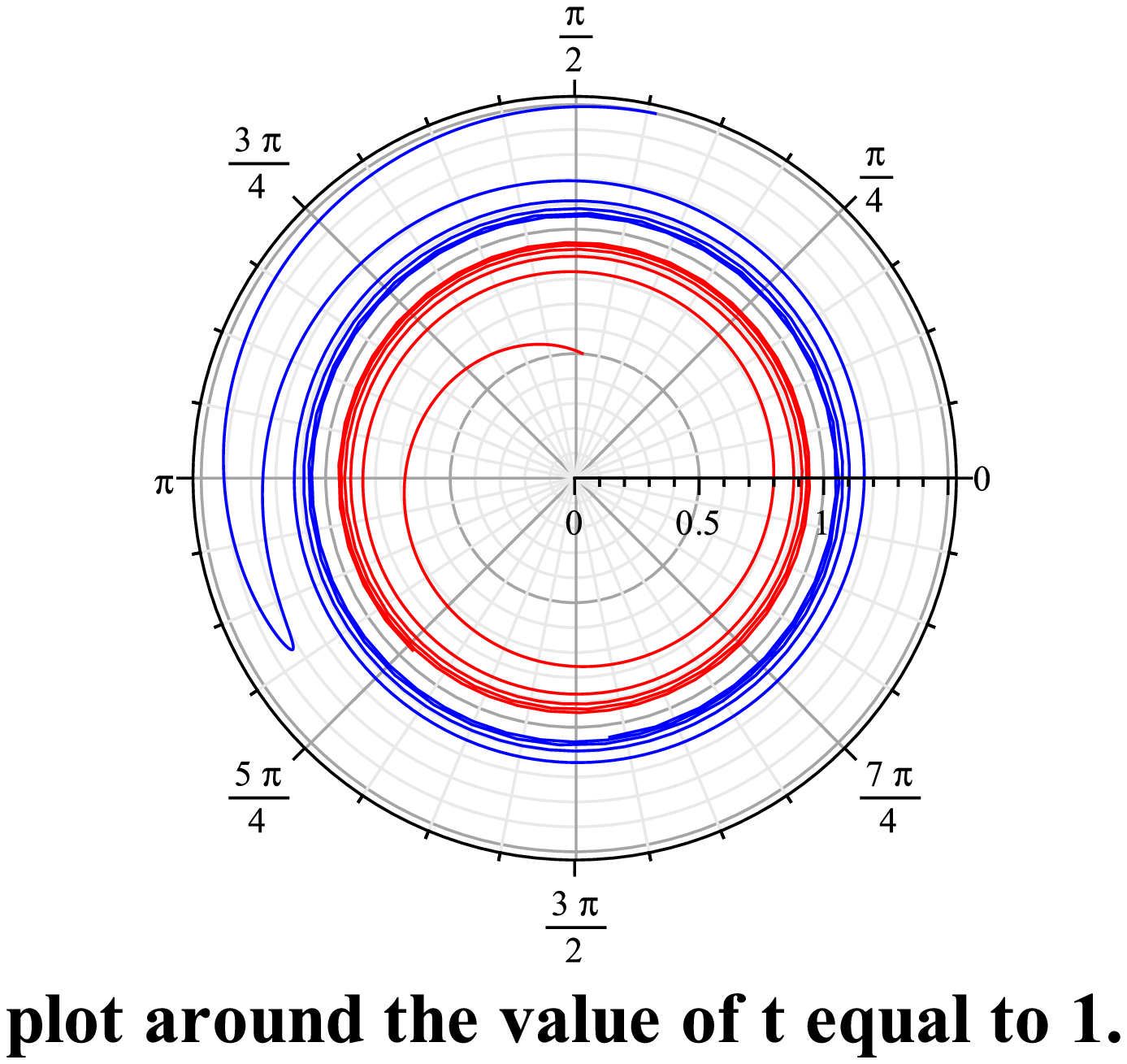,width=5.5cm,height=4cm} }\\
 { \psfig{figure=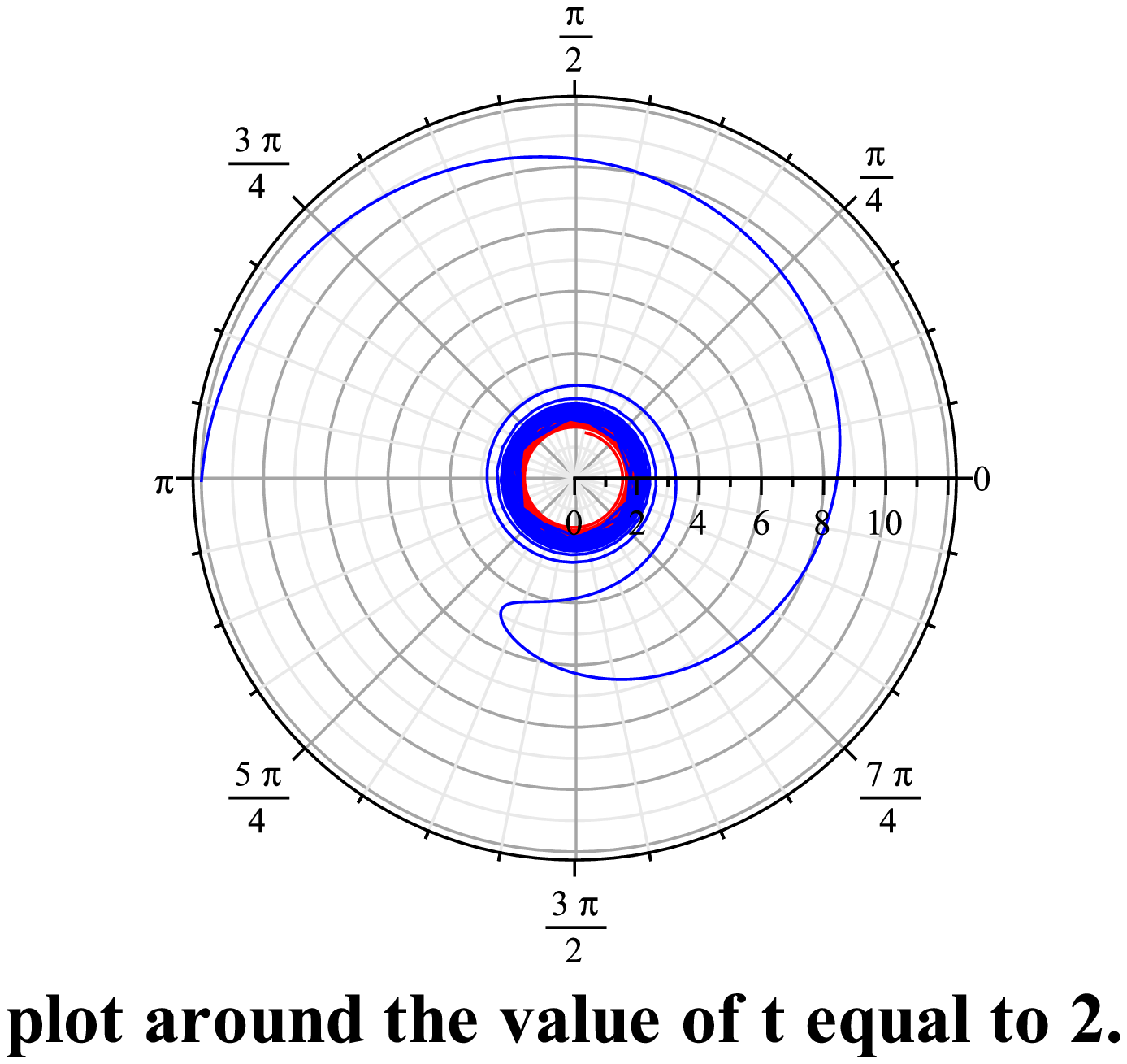,width=5.5cm,height=4cm} }&
 { \psfig{figure=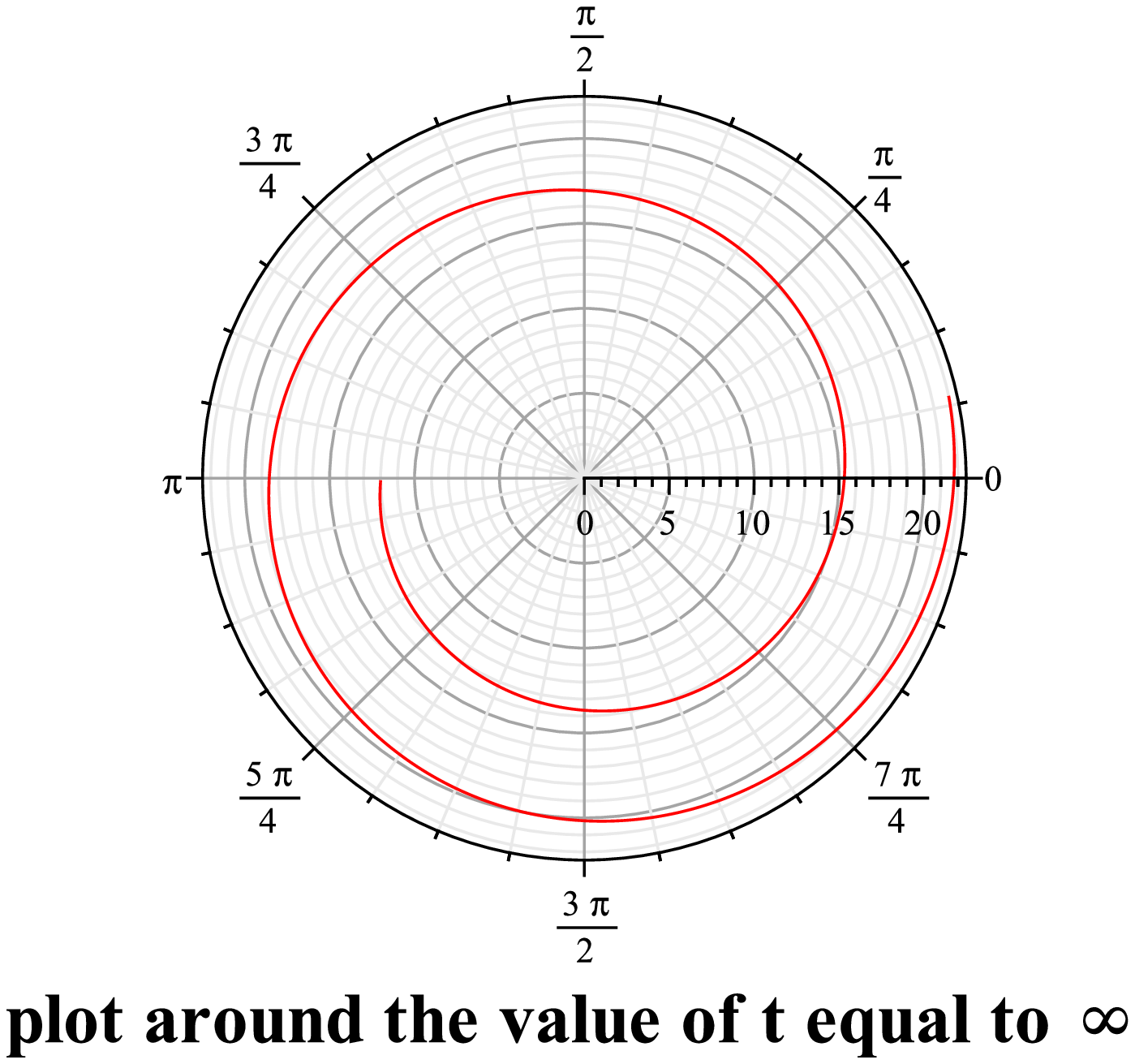,width=5.5cm,height=4cm} }&
  \end{array}
$$
\caption{ $\varphi_6(t)$}
\label{15}
\end{figure}
\section{Conclusions/Further Work}

In this paper we have presented several theoretical results and an algorithm for properly plotting curves parametrized by rational functions in polar form. Our results allow to algorithmically identify phenomena which are typical of these curves, like the existence of infinitely many self-intersections, spiral branches, limit points or limit circles. Furthermore, the algorithm
has been implemented in Maple 15, and provides good results. Natural extensions of the study here are space curves which are rational in spherical or cylindrical coordinates, curves which are algebraic, although non-necessarily rational, in polar coordinates (i.e. fulfilling $h(r,\theta)=0$, with $h$ algebraic), and similar phenomena for the case of surfaces. It would be also interesting to analyze the curves defined by (implicit) expressions
of the type $f(r,\mbox{sin}(\theta),\mbox{cos}(\theta))=0$, where $f$ is algebraic, since this class contains, and in fact extends, the class of algebraic curves; also, it includes the important subclass of curves defined by equations $r^n=g(\theta)$, with $g(\theta)$ a rational function, which often appear in Geometry and Physics. Some of these questions will be explored in the future.

\end{document}